\documentclass[twocolumn]{aastex631}

\usepackage{multirow}
\usepackage{color}
\usepackage{float}
\usepackage{tabularx}
\makeatother
\makeatletter
\usepackage{comment}
\usepackage{natbib}
\usepackage{amsmath}
\usepackage{booktabs}
\usepackage{array}
\usepackage{hyperref}
\hypersetup{colorlinks=true, citecolor=blue, filecolor=magenta,
 urlcolor=cyan, linkcolor=blue}

\begin{document}

\defcitealias{Zavala2021}{Z21}
\defcitealias{Casey2021}{C21}

\title{The Extended Mapping Obscuration to Reionization with ALMA (Ex-MORA) Survey: 5$\sigma$ Source Catalog and Redshift Distribution }

\author[0000-0002-7530-8857]{Arianna S. Long}\altaffiliation{E-mail: aslong@uw.edu}
\affiliation{Department of Astronomy, The University of Washington, Seattle, WA USA}

\author[0000-0002-0930-6466]{Caitlin M. Casey}
\affiliation{Department of Astronomy, The University of Texas at Austin, Austin, TX, USA}

\author[0000-0002-6149-8178]{Jed McKinney}\altaffiliation{NASA Hubble Fellow}
\affiliation{Department of Astronomy, The University of Texas at Austin, Austin, TX, USA}

\author[0000-0002-7051-1100]{Jorge A. Zavala}
\affiliation{National Astronomical Observatory of Japan, 2-21-1 Osawa, Mitaka, Tokyo 181-8588, Japan}

\author[0000-0002-0930-6466]{Hollis B. Akins}
\affiliation{Department of Astronomy, The University of Texas at Austin, Austin, TX, USA}

\author[0000-0003-3881-1397]{Olivia R. Cooper}\altaffiliation{NSF Graduate Fellow}
\affiliation{Department of Astronomy, The University of Texas at Austin, Austin, TX, USA}

\author[0000-0003-3216-7190]{Erini L. Lambrides}
\altaffiliation{NASA Postdoctoral Fellow}
\affiliation{NASA-Goddard Space Flight Center, Code 662, Greenbelt, MD, 20771, USA}

\author[0000-0002-3560-8599]{Maximilien Franco}
\affiliation{Department of Astronomy, The University of Texas at Austin, Austin, TX, USA}

\author[0000-0002-6290-3198]{Manuel Aravena}
\affiliation{Instituto de Estudios Astrofísicos, Facultad de Ingeniería y Ciencias, Universidad Diego Portales, Av. Ejército 441, Santiago,
Chile}

\author[0000-0002-3915-2015]{Matthieu Bethermin}
\affiliation{Université de Strasbourg, CNRS, Observatoire astronomique de
Strasbourg, UMR 7550, 67000 Strasbourg, France}
\affiliation{Aix Marseille Univ, CNRS, CNES, LAM, Marseille, France}

\author[0000-0001-8183-1460]{Karina Caputi}
\affiliation{Kapteyn Astronomical Institute, University of Groningen, P.O. Box 800, 9700AV Groningen, The Netherlands}

\author[0000-0002-6184-9097]{Jaclyn B. Champagne}
\affiliation{Steward Observatory, University of Arizona, 933 N Cherry Ave, Tucson, AZ 85721, USA}

\author[0000-0002-9548-5033]{D. L. Clements}
\affiliation{Astrophysics Group, Imperial College London, Blackett Laboratory, Prince Consort Road, London SW7 2AZ, UK}

\author[0000-0001-9759-4797]{Elisabete da Cunha}
\affiliation{International Centre for Radio Astronomy Research, University of Western Australia, 35 Stirling Hwy, Crawley 26WA 6009, Australia}

\author[0000-0002-9382-9832]{Andreas L. Faisst}
\affiliation{Caltech/IPAC, MS 314-6, 1200 E. California Blvd. Pasadena, CA 91125, USA}

\author[0000-0002-8008-9871]{Fabrizio Gentile}
\affiliation{University of Bologna, Department of Physics and Astronomy (DIFA), Via Gobetti 93/2, I-40129, Bologna, Italy}
\affiliation{INAF—Osservatorio di Astrofisica e Scienza dello Spazio, via Gobetti 93/3-I-40129, Bologna, Italy}

\author[0000-0001-6586-8845]{Jacqueline Hodge}
\affiliation{Leiden Observatory, Leiden University, P.O. Box 9513, 2300 RA Leiden, the Netherlands}

\author[0000-0003-2475-124X]{Allison W. S. Man}
\affiliation{Department of Physics \& Astronomy, University of British Columbia, 6224 Agricultural Road, Vancouver BC, V6T 1Z1, Canada}

\author[0000-0003-0415-0121]{Sinclaire M. Manning}
\altaffiliation{NASA Hubble Fellow}
\affiliation{Department of Astronomy, University of Massachusetts, Amherst, MA 01003, USA}

\author[0000-0002-1233-9998]{David B. Sanders}
\affiliation{Institute for Astronomy, University of Hawai’i at Manoa, 2680 Woodlawn Drive, Honolulu, HI 96822, USA}

\author[0000-0003-4352-2063]{Margherita Talia}
\affiliation{University of Bologna, Department of Physics and Astronomy (DIFA), Via Gobetti 93/2, I-40129, Bologna, Italy}
\affiliation{INAF—Osservatorio di Astrofisica e Scienza dello Spazio, via Gobetti 93/3-I-40129, Bologna, Italy}

\author[0000-0001-7568-6412]{Ezequiel Treister}
\affiliation{Instituto de Astrofísica, Facultad de Física, Pontificia Universidad Católica de Chile, Casilla 306, Santiago 22, Chile}

\author[0000-0003-2680-005X]{Gabriel Brammer}
\affiliation{The Cosmic Dawn Center, Jagtvej 155A, 2200 København N, Denmark}

\author[0000-0002-5059-6848]{Marcella Brusa}
\affiliation{Dipartimento di Fisica e Astronomia, Università di Bologna, Via Gobetti 93/2, 40129 Bologna, Italy}

\author[0000-0001-8519-1130]{Steven L. Finkelstein}
\affiliation{Department of Astronomy, The University of Texas at Austin, Austin, TX, USA}

\author[0000-0001-7201-5066]{Seiji Fujimoto}\altaffiliation{Hubble Fellow}
\affiliation{Department of Astronomy, The University of Texas at Austin, Austin, TX, USA}

\author[0000-0003-4073-3236]{Christopher C. Hayward}
\affiliation{Center for Computational Astrophysics, Flatiron Institute, 162 Fifth Avenue, New York, NY 10010, USA}

\author[0000-0002-7303-4397]{Olivier Ilbert}
\affiliation{Aix Marseille Univ, CNRS, CNES, LAM, Marseille, France  }

\author[0000-0002-3405-5646]{Jean-Baptiste Jolly}
\affiliation{Max-Planck-Institut für Extraterrestrische Physik (MPE), Giessenbachstraße 1, D-85748 Garching, Germany}

\author[0000-0001-9187-3605]{Jeyhan S. Kartaltepe}
\affiliation{Laboratory for Multiwavelength Astrophysics, School of Physics and Astronomy, Rochester Institute of Technology, Rochester, NY USA}

\author{Kirsten Knudsen}
\affiliation{Department of Space, Earth and Environment, Chalmers University of Technology, SE-412 96 Gothenburg, Sweden}

\author[0000-0002-6610-2048]{Anton M. Koekemoer}
\affiliation{Space Telescope Science Institute, 3700 San Martin Dr., Baltimore, MD 21218, USA} 

\author[0000-0001-9773-7479]{Daizhong Liu}
\affiliation{Purple Mountain Observatory, Chinese Academy of Sciences, 10 Yuanhua Road, Nanjing 210023, China}

\author[0000-0002-4872-2294]{Georgios Magdis}
\affiliation{The Cosmic Dawn Center, Jagtvej 155A, 2200 København N, Denmark}
\affiliation{DTU-Space, Technical University of Denmark, Elektrovej 327, 2800, Kgs. Lyngby, Denmark}
\affiliation{Niels Bohr Institute, University of Copenhagen, Jagtvej 128, DK-2200, Copenhagen, Denmark}

\author[0000-0002-9489-7765]{Henry Joy McCracken}
\affiliation{Institut d’Astrophysique de Paris, UMR 7095, CNRS, and Sorbonne Université, 98 bis boulevard Arago, F-75014 Paris, France}

\author[0000-0002-4485-8549]{Jason Rhodes}
\affiliation{Jet Propulsion Laboratory, California Institute of Technology, Pasadena, CA, USA}

\author[0000-0002-4271-0364]{Brant E. Robertson}
\affiliation{Department of Astronomy and Astrophysics, University of California, Santa Cruz, Santa Cruz, CA USA}

\author[0000-0002-0438-3323]{Nick Scoville}
\affiliation{Astronomy Dept., California Institute of Technology, 1200 E. California Blvd, Pasadena, CA, USA}

\author[0000-0002-5496-4118]{Kartik Sheth}
\affiliation{NASA Headquarters, 300 Hidden Figures Way, SE, Mary W. Jackson NASA HQ Building, Washington, DC 20546, USA}

\author[0000-0002-3893-8614]{Vernesa Smolcic}
\affiliation{Department of Physics, University of Zagreb, Bijenicka cesta 32, 10002 Zagreb, Croatia}

\author[0000-0003-3256-5615]{Justin Spilker}
\affiliation{Department of Physics and Astronomy \& George P. and Cynthia Woods Mitchell Institute for Fundamental
Physics and Astronomy, Texas A\&M University, College Station, TX, USA}

\author[0000-0003-2247-3741]{Yoshiaki Taniguchi}
\affiliation{The Open University of Japan, 2-11, Wakaba, Mihama-ku, Chiba 261-8586, Japan}

\author[0000-0003-3631-7176]{Sune Toft}
\affiliation{The Cosmic Dawn Center, Jagtvej 155A, 2200 København N, Denmark}
\affiliation{Niels Bohr Institute, University of Copenhagen, Jagtvej 128, DK-2200, Copenhagen, Denmark}

\author[0000-0002-0745-9792]{C. Megan Urry}
\affiliation{Physics Department and Yale Center for Astronomy \& Astrophysics, Yale University, PO Box 208120, CT 06520-8120, USA}

\author[0000-0001-7095-7543]{Min Yun}
\affiliation{Department of Astronomy, University of Massachusetts, Amherst, MA 01003, USA}

\begin{abstract}

One of the greatest challenges in galaxy evolution over the last decade has been constraining the prevalence of heavily dust-obscured galaxies in the early Universe. At $z>3$, these galaxies are increasingly rare, and difficult to identify as they are interspersed among the more numerous dust-obscured galaxy population at $z=1-3$, making efforts to secure confident spectroscopic redshifts expensive, and sometimes unsuccessful. In this work, we present the Extended Mapping Obscuration to Reionization with ALMA (Ex-MORA) Survey -- a 2\,mm blank-field survey in the COSMOS-Web field, and the largest ever ALMA blank-field survey to-date covering 577\,arcmin$^2$. Ex-MORA is an expansion of the MORA survey designed to identify primarily $z>3$ dusty, star-forming galaxies while simultaneously filtering out the more numerous $z<3$ population by leveraging the very negative $K$-correction at observed-frame 2\,mm. We identify 37 significant ($>$\,5$\sigma$) sources, 33 of which are robust thermal dust emitters. We measure a median redshift of $\langle z \rangle = 3.6^{+0.1}_{-0.2}$, with two-thirds of the sample at $z>3$, and just under half at $z>4$, demonstrating the overall success of the 2\,mm--selection technique. We find an integrated $z>3$ volume density of Ex-MORA sources of $\sim1-3\times10^{-5}$\,Mpc$^{-3}$, consistent with other surveys of infrared luminous galaxies at similar epochs. We also find that techniques using rest-frame optical emission (or lack thereof) to identify $z>3$ heavily dust-obscured galaxies miss at least half of Ex-MORA galaxies. This supports the idea that the dusty galaxy population is heterogeneous, and that synergies across several major observatories spanning multiple energy regimes are critical to understanding their formation and evolution at $z>3$.



\end{abstract}


\section{Introduction} \label{sec:intro}
Dusty, star-forming galaxies (DSFGs) are a class of extraordinarily infrared luminous (L$_\mathrm{IR} > 10^{12}$\,L$_\odot$) galaxies that are most abundant during the peak of stellar growth in the Universe \citep[$z \sim 1-3$,][]{Smail1997, Blain2002, casey2014dusty}. At the peak of their population prevalence, DSFGs -- also known as submillimeter galaxies -- are believed to be the most massive (M$_\star \sim 10^{11}$\,M$_\odot$) and most prolific star-forming galaxies, with star formation rates (SFRs) exceeding hundreds of solar masses per year \citep{casey2014dusty,hodgeanddacunha}. Understanding this extreme population is critical for constraining many areas of astrophysics, including extreme stellar feedback and outflow models \citep{Hayward2021, Heintz2022, Bassini2023}, rapid metal enrichment of the cosmos \citep{madau14, Shen2022, Birkin2023}, the formation of the first massive quiescent galaxies \citep{toft14, Valentino2020, Long2023}, and the collapse of dark matter halos in overdense environments at high-$z$ \citep{hickox2012, casey16, lewis18, Long2020, Lim2021}. This is why, over the last two decades, it has become increasingly important to chart the prevalence, properties, and progression of DSFGs across cosmic time. 

There is still mystery, however, surrounding the true abundance of luminous DSFGs in the first 2\,Gyr of the cosmos, with up to two orders of magnitude in discrepancy across various studies at $z\gtrsim3$ \citep[see e.g.,][for summaries of current estimates across the literature]{Long2023, Valentino2023}. This is largely because identifying $z>3$ DSFGs is difficult: their extreme dust-obscuration makes their starlight faint, or perhaps entirely undetectable \citep[e.g.,][]{walter12, Williams2019, Manning2022, McKinney2023b, Gentile2024}, at rest-frame ultraviolet and optical wavelengths such that the typical emission lines used to secure redshifts of star-forming galaxies are attenuated beyond detection \citep[though new JWST observations are starting to prove effective at detecting these faint lines, e.g.,][]{Herard-Demanche2023, Xiao2023}. Photometric redshift estimates based on the shapes of their far-IR/sub-millimeter spectral energy distributions (SEDs) are highly uncertain and often span a wide range of potential redshifts ($z\sim1-12$) due to temperature-redshift degeneracies \citep{Casey2020}; and far-IR spectroscopic scans are expensive as they often require integration times on the order of $\sim1.5-3$\,hrs \textit{per target} due to smaller fields of view, instrument sensitivity limitations, and the need to cover a wide spectral range (e.g. \citealt[][]{Gentile2024b}, though gravitationally lensed sources can instead take mere minutes, \citealt{Reuter2020}). 

\citet{Casey2018a, Casey2018b} developed a backward galaxy evolution model of DSFGs based on empirical data that demonstrated how -- with the appropriate depths -- long wavelength observations ($\lambda_\mathrm{observed} \sim 2-3$\,mm) can be an efficient way to detect DSFGs at $z>3$, while simultaneously filtering out the more numerous $z<3$ DSFG population. Early results leveraging this method prove promising, with several 2\,mm \citep{Magnelli2019, Casey2021, Cooper2022, Cowie2023} and 3\,mm \citep{zavala18, Williams2019} programs identifying high fractions of $z>3$ DSFGs in comparison to surveys at $\lambda_\mathrm{observed} \lesssim 1$\,mm. This is the basis for the original Mapping Obscuration to Reionization with ALMA (MORA) Survey \citep{Zavala2021, Casey2021}, a contiguous blank-field mapping survey at 2\,mm in the Cosmic Evolution Survey (COSMOS) field \citep{Capak2007, Koekemoer2007, Scoville2007}. The original survey covered 184\,arcmin$^2$ and discovered 12 robust sources, with over 70\% of the sample estimated to lie at $z>3$. 

\begin{figure*}[ht]
    \centering
    \includegraphics[width = 1\textwidth]{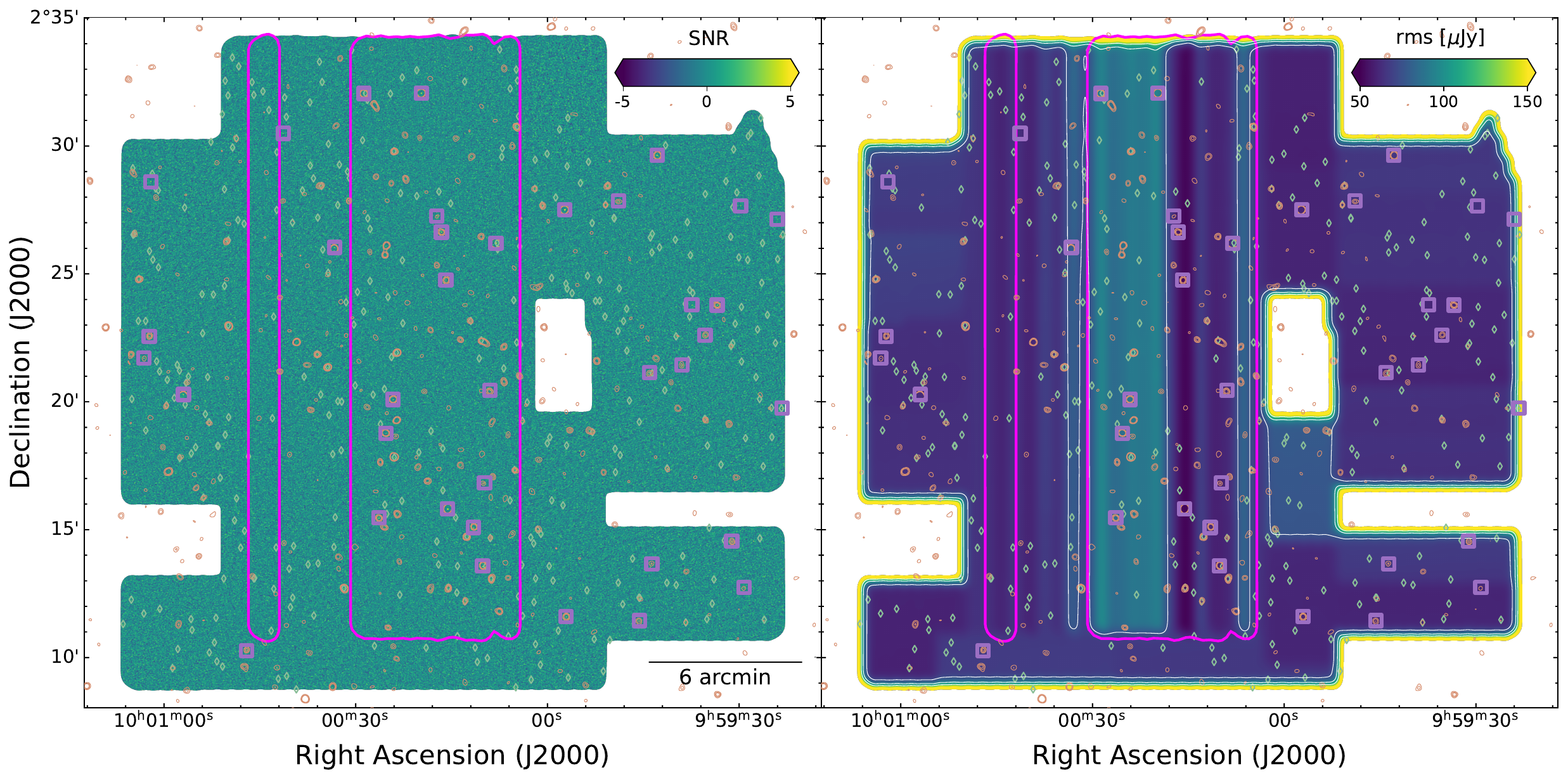}
    \caption{Signal-to-noise ratio (SNR, left) and rms (right) map of the Ex-MORA mosaic. The original MORA survey presented in \citetalias{Zavala2021} and \citetalias{Casey2021} is outlined in magenta. Sources at $>5\sigma$ significance are marked by purple boxes, while sources with $4\sigma <$\,SNR\,$<5\sigma$ are marked by green diamonds. Contours from the SCUBA-2 850\,$\mu$m Survey of the COSMOS Field \citep{Simpson2019} are overlaid in orange at levels of 3.5$\sigma$, 5$\sigma$, and 6.5$\sigma$ significance. White contours on the rms map increase from 50 to 150\,$\mu$Jy\,beam$^{-1}$ in steps of 25\,$\mu$Jy\,beam$^{-1}$. }
    \label{fig:mora-map}
\end{figure*}

\begin{figure}
    \centering
    \includegraphics[width=0.45\textwidth]{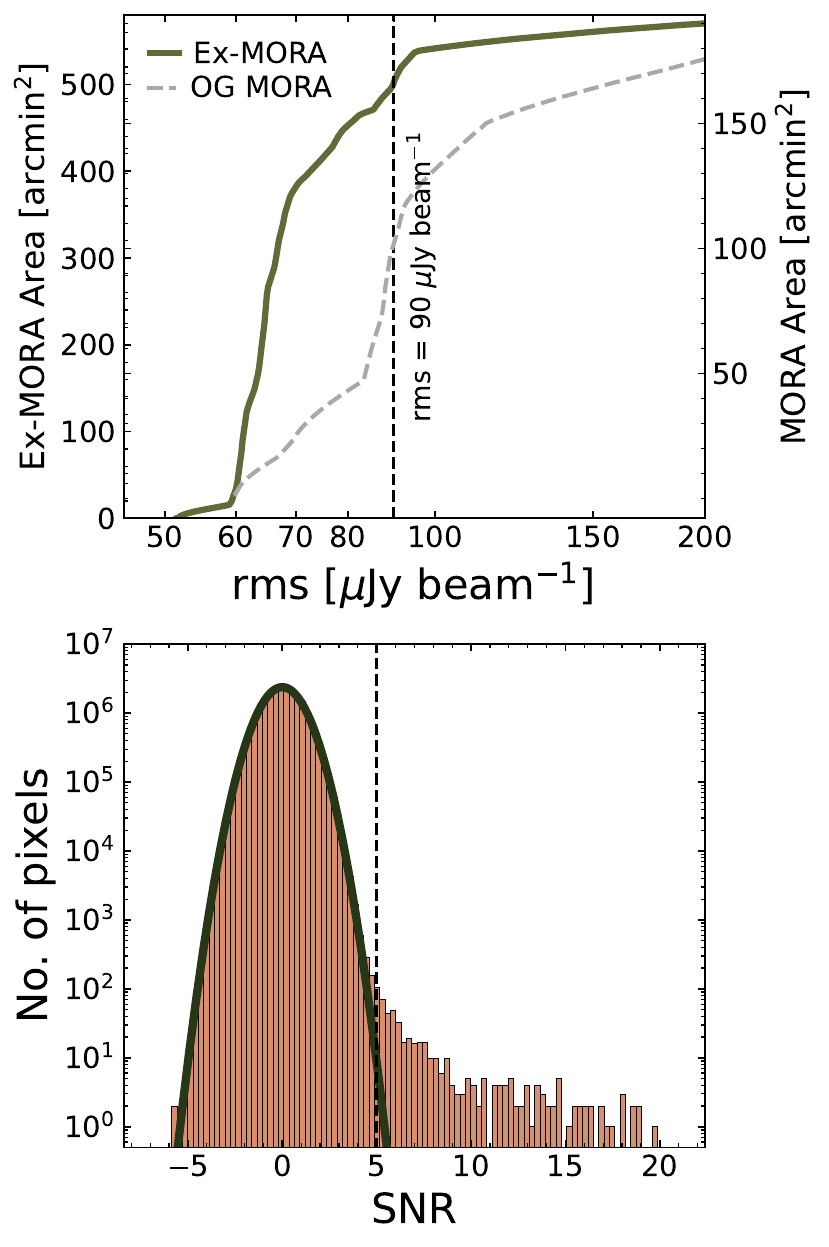}
    \caption{On the top, we show the cumulative distribution of the 1$\sigma$ survey depth for Ex-MORA (green solid) and original (OG) MORA (grey dashed). The majority of the Ex-MORA survey (90\%) has depths at or below 90\,$\mu$Jy\,beam$^{-1}$; this is only $\sim60\%$ for OG MORA. On the bottom, we show the histogram of pixel values in the signal-to-noise maps for Ex-MORA. The histogram is well fit by a standard normal (Gaussian) distribution with $\sigma = 1$ (overlaid solid line), confirming Gaussian properties of the map's noise. We see a divergence above $5\sigma$ due to the presence of positive source peaks. }
    \label{fig:noise}
\end{figure}

In this Paper, we present an update to the original MORA Survey: the \textit{extended} or Ex-MORA Survey. Ex-MORA is a factor of $\sim3\times$ larger than the original MORA survey, covering a total of 0.16\,deg$^2$ over the COSMOS-Web Cycle 1 JWST Survey area \citep{cosmos-web}. In Section \ref{sec:EM survey}, we present the survey design, reduction, and map characteristics. In Section \ref{sec:sample selection}, we present the 2\,mm-bright sources detected at $\ge5\sigma$ in this survey, and their respective redshifts. Then, in Section \ref{sec:results}, we present results on the cumulative number density of 2\,mm sources, their redshift distributions, and highlight some interesting sample outliers, then provide our conclusions in Section \ref{sec:conclusions}. Where relevant, we assume a standard $\Lambda$CDM cosmology with \textit{Planck}-measured parameters \citep{PlanckCollaboration2020}, with H$_0 = 67.4$\,km\,s$^{-1}$\,Mpc$^{-1}$, and a Chabrier initial mass function \citep{chabrier}. 

\begin{deluxetable*}{lcccccc}[ht]
\centering
\tablecaption{Ex-MORA Survey Observation Characteristics}
 \tablehead{
\colhead{SB Name} & \colhead{R.A.} & \colhead{Dec} & \colhead{PWV} & \colhead{On Source Time} & \colhead{rms} & \colhead{Synth. Beam Size} \\
\colhead{} & \colhead{} & \colhead{} & \colhead{(mm)} & \colhead{(min)} & \colhead{($\mu$Jy\,beam$^{-1}$)} & \colhead{}}
 \startdata
    POS-01 & 10:00:48.8 & $+$02:22:30.0 & 0.91 & 37.95 & 87 & $1\farcs71 \times 1\farcs43$\\
    POS-02 & 10:00:46.6 & $+$02:22:30.0 & 1.49 & 37.95 & 89 & $1\farcs59 \times 1\farcs44$\\
    POS-04 & 10:00:42.1 & $+$02:22:30.0 & 1.03 & 37.97 & 91 & $1\farcs78 \times 1\farcs47$\\
    POS-05 & 10:00:39.9 & $+$02:22:30.0 & 1.37 & 37.98 & 85 & $1\farcs65 \times 1\farcs40$\\
    POS-06 & 10:00:37.7 & $+$02:22:30.0 & 1.28 & 37.97 & 94 & $1\farcs82 \times 1\farcs35$\\
    POS-07 & 10:00:35.4 & $+$02:22:30.0 & 1.21 & 37.97 & 91 & $1\farcs64 \times 1\farcs43$\\
    POS-08 & 10:00:33.2 & $+$02:22:30.0 & 1.43 & 38.00 & 112 & $2\farcs06 \times 1\farcs28$\\
    POS-09 & 10:00:31.0 & $+$02:22:30.0 & 1.49 & 38.00 & 93 & $1\farcs99 \times 1\farcs28$\\
    POS-16 & 10:00:15.4 & $+$02:22:30.0 & 0.87 & 37.98 & 78 & $1\farcs65 \times 1\farcs40$\\
    POS-19 & 10:00:08.7 &$+$02:22:30.0  & 1.30 & 38.00 & 83 & $1\farcs78 \times 1\farcs37$\\
    POS-21 & 10:00:04.2 & $+$02:22:30.0 & 1.37 & 37.98 & 89 & $1\farcs64 \times 1\farcs52$\\
    EPOS-02 & 09:59:39.3 & $+$02:31:22.3 & 0.97 & 35.28 & 86 & $1\farcs75 \times 1\farcs49$\\
    EPOS-03 & 09:59:38.1 & $+$02:29:12.9 & 1.45 & 39.85 & 88 & $1\farcs95 \times 1\farcs39$\\
    EPOS-04 & 09:59:38.0 & $+$02:27:14.8 & 1.47 & 39.82 & 87 & $1\farcs74 \times 1\farcs46$\\
    EPOS-05 & 09:59:38.5 & $+$02:25:20.1 & 1.48 & 37.72 & 90 & $1\farcs64 \times 1\farcs51$\\
    EPOS-06 & 09:59:38.0 & $+$02:23:33.0 & 0.65 & 39.83 & 87 & $1\farcs92 \times 1\farcs39$\\
    EPOS-07 & 09:59:38.0 & $+$02:21:37.1 & 0.61 & 39.82 & 84 & $1\farcs69 \times 1\farcs48$\\
    EPOS-08 & 09:59:38.0 & $+$02:19:41.2 & 1.47 & 39.83 & 89 & $1\farcs72 \times 1\farcs37$\\
    EPOS-09 & 09:59:38.0 & $+$02:17:45.2 & 0.69 & 39.85 & 90 & $1\farcs91 \times 1\farcs30$\\
    EPOS-10 & 09:59:38.0 & $+$02:15:49.3 & 1.01 & 39.85 & 87 & $1\farcs66 \times 1\farcs52$\\
    EPOS-11 & 09:59:38.0 & $+$02:13:53.4 & 1.58 & 39.83 & 94 & $1\farcs76 \times 1\farcs48$\\
    EPOS-12 & 09:59:38.0 & $+$02:11:57.5 & 0.66 & 39.83 & 86 & $1\farcs59 \times 1\farcs43$\\
    EPOS-14 & 09:59:57.5 & $+$02:31:35.8 & 0.39 & 39.83 & 79 & $2\farcs28 \times 2\farcs03$\\
    EPOS-15 & 09:59:57.5 & $+$02:26:46.0 & 0.56 & 39.87 & 80 & $1\farcs58 \times 1\farcs41$\\
    EPOS-17 & 09:59:57.5 & $+$02:16:56.7 & 1.82 & 39.83 & 105 & $2\farcs16 \times 1\farcs29$\\
    EPOS-18 & 09:59:57.5 & $+$02:12:07.0 & 0.84 & 39.82 & 85 & $1\farcs74 \times 1\farcs34$\\
    EPOS-19 & 10:00:12.4 & $+$02:09:55.9 & 1.02 & 38.75 & 94 & $1\farcs72 \times 1\farcs36$\\
    EPOS-20 & 10:00:40.5 & $+$02:10:01.6 & 1.76 & 39.82 & 94 & $1\farcs77 \times 1\farcs37$\\
    EPOS-21 & 10:00:58.7 & $+$02:11:09.0 & 0.60 & 38.22 & 84 & $1\farcs64 \times 1\farcs51$\\
    EPOS-23 & 10:00:57.7 & $+$02:18:01.7 & 1.65 & 39.85 & 88 & $1\farcs59 \times 1\farcs44$\\
    EPOS-24 & 10:00:57.8 & $+$02:21:27.2 & 1.48 & 39.33 & 111 & $1\farcs57 \times 1\farcs42$\\
    EPOS-25 & 10:00:57.8 & $+$02:24:52.8 & 1.55 & 39.85 & 91 & $1\farcs93 \times 1\farcs34$\\
    EPOS-26 & 10:00:57.8 & $+$02:28:18.5 & 0.97 & 39.28 & 89 & $2\farcs03 \times 1\farcs45$\\
 \hline
 OG MORA & ... & ... & ... & ... & $60-90$ & $1\farcs83 \times 1\farcs43$\\
 \hline
 Combined Mosaic & ... & ... & ... & ... & $50-260$ & $1\farcs68\times1\farcs44$\\
 \enddata 
\end{deluxetable*} \label{table:obs}

\section{The Extended Mapping Obscuration to Reionization with ALMA (Ex-MORA) Survey} \label{sec:EM survey}

The Ex-MORA Survey (ALMA Project Code 2021.1.00225.S, PI: C.M. Casey) is an extension to the MORA survey presented in \citet[][hereafter Z21]{Zavala2021} and \citet[][hereafter C21]{Casey2021}. This updated survey was designed to extend the original MORA 2\,mm blank-field coverage of the COSMOS-Web field \citep{cosmos-web} from 0.05\,deg$^2$ to a contiguous 0.2\,deg$^2$. Though A-rated, the project was only partially observed, resulting in a large, contiguous mosaic of $\sim$\,577\,arcmin$^2$ or 0.16\,deg$^2$ centered on $\alpha \approx$\,10:00:15, $\delta \approx$\,+02:21:32 (shown in Figure \ref{fig:mora-map}). This is roughly a factor of three larger than the original MORA survey area (which is subsumed in the Ex-MORA map), covering roughly a third of the COSMOS-Web field. The deepest portion of the mosaic has an rms of $\sigma_\mathrm{2\,mm} = 51$\,$\mu$Jy\,beam$^{-1}$, with $\sim$90\% of the map ($\sim$\,510\,arcmin$^2$) at or below the proposed map depth of 90\,$\mu$Jy\,beam$^{-1}$ (Figure \ref{fig:mora-map}).

Observations took place during ALMA Cycle 8, from April 2022 to January 2023 in a nominal C43-2 configuration in band 4 centered on a 147\,GHz local oscillator frequency tuning. We note that the original MORA survey was proposed to be observed over two different tunings (one centered on 147\,GHz and one on 139\,GHz) to optimize potential detections of emission lines at $z\sim2.5$; though original MORA was not completed, and most of the observed SBs were centered on 147\,GHz. Thus, Ex-MORA was requested at only the one 147\,GHz tuning in order to reduce overheads and to enable efficient mosaicking with original MORA survey. The 33 Ex-MORA scheduling blocks (SBs) were spatially distributed in a similar fashion as the original MORA survey. The unfinished SBs from the original MORA survey take the form of elongated ``breadstick'' mosaics of $\sim149$ pointings, and each of those pointings are 19$\farcs$3 offset in R.A. Newly designed Ex-MORA SBs are more square in their on-sky shape, and distributed around the perimeter of MORA, but fully aligned with the original MORA grid (following Nyquist spacing). The average phase center of each SB is reported in Table \ref{table:obs}, along with individual noise and precipitable water vapor (PWV) properties of each SB. 

The mosaic was reduced and imaged in the same manner as presented in \citetalias{Zavala2021} and \citetalias{Casey2021}, including the following key parameters: we imaged these data using natural weighting, with a robust value $=2$ to optimize source signal-to-noise ratios, set the cell / pixel size to 0.3 arcseconds (which is well below the synthesized beam size, see below), and averaged the maps over time bins of 10s. Observations were taken under an average and favorable precipitable water vapor of 1.10\,mm. There are a total of 4851 individual pointings in the Ex-MORA mosaic; with $142-150$ pointings per SB and an average of 22 seconds per pointing, this combines to a total of 29.4\,hrs (including overheads and calibrations). Measurement sets for the individual SBs were concatenated together using the CASA \texttt{concat} command, then reduced collectively as a mosaic. The combined mosaic has a typical synthesized beam size of $\theta_\mathrm{FWHM} \approx 1\farcs68 \times 1\farcs44$, which is similar to the average beam sizes across individual SBs (see Table \ref{table:obs}). This beam size translates to physical scales of $\sim 13 \times 11$\,kpc at $z=3$, which is well beyond the observed physical extent of dust emission in galaxies at these epochs \citep{hodge16, Gillman2023}. Thus, it is unlikely that sources are resolved out of the map, or spatially resolved in general. Therefore, we adopt source positions and flux densities corresponding to the point of their peak signal-to-noise by identifying all of the $>$5$\sigma$ pixels using a ``region grow'' algorithm \citep{Casey2018b, Cooper2022}. As shown in Figure \ref{fig:noise}, the SNRs of the map pixels are well represented by a standard normal centered about zero with $\sigma = 1$, above which positive source peaks diverge from the distribution, demonstrating that the SNR\,$\ge5$\,$\sigma$ threshold is an ideal threshold for low source contamination. 

A series of thorough tests on survey completeness, false detection rates, and flux boosting effects were carried out in \citetalias{Zavala2021}. We repeat these tests and calculate the corresponding corrections in the same fashion, and summarize briefly here. To quantify the false detection rate, we create mock noise maps that are the same size as the Ex-MORA map, with Gaussian noise distributed randomly. We then convolve this map with the average synthesized beam listed above, and re-normalize the convolved map to a standard normal distribution to represent a pure noise (source--free) map. We then count the number of serendipitous peaks at or above 5$\sigma$ significance. We repeat this process 100 times, and calculate an average of $0.7^{+0.7}_{-0.5}$ of false detections in the Ex-MORA map, with an average SNR of $5.3\pm 0.2$. This is in good agreement with estimates in \citetalias{Zavala2021} and \citetalias{Casey2021}, as well as our own catalog for which we identify one likely spurious source at the same SNR as the false detection estimates (see next Section). Survey completeness is calculated by inserting artificial sources at random positions around the mosaic, then using the same peak finding algorithm to recover them. This process is repeated 100 times, calculating the ratio between recovered and inserted sources on each iteration. We calculate the average completeness as a function of SNR, finding $\sim80\%$ survey completeness for sources with SNRs $\ge 6\sigma$, while sources at $5-6\sigma$ are most affected with source completeness estimated at $\sim50-70$\%. We use these same simulations to estimate flux-boosting effects by taking the ratio of the input and recovered flux, and find these effects to be minimal ($\sim5\%$). These corrections are folded into the resulting on-sky number density calculation, as described in Section \ref{sec:onskydens}.

\section{The 2\,mm Selected Sample} \label{sec:sample selection}

In this section, we describe the Ex-MORA source catalog, and the ancillary data used to identify source counterparts and estimate their respective redshifts. The goal of this work is to update statistics on the efficiency of the 2\,mm selection technique in identifying $z\gtrsim3$ massive, dust-obscured galaxies. Detailed analysis of the physical properties, morphologies, and environments of these sources will be presented in forthcoming papers. 

\subsection{Ancillary Data}

The Ex-MORA Survey benefits from complete coverage with the COSMOS-Web Survey (GO \#1727, PIs: Casey \& Kartaltepe; \citealt{cosmos-web}), a 255\,hr imaging program with contiguous coverage over 0.54\,deg$^2$ in four JWST NIRCam filters (F115W, F150W, F277W, and F444W), and parallel MIRI imaging in the F770W band. Details on the survey design can be found in \citet{cosmos-web}, and details on data reduction will be described in full in M. Franco et al. (2024, \textit{in prep.}). In summary, JWST NIRCam data for the epoch used in this analysis are reduced with the JWST Calibration Pipeline version 1.12.1 \citep{Bushouse2023}, with CRDS pmap-1170, corresponding to NIRCam instrument mapping imap-0273. We implement additional custom modifications used in other ERS and Cycle 1 surveys \citep[e.g., ][]{Bagley2023}, including subtraction of 1/$f$ and background noise. NIRCam images achieve a 30\,mas/pixel resolution, with astrometry aligned to the COSMOS2020 catalog \citep{Weaver2022}, which is anchored to Gaia-EDR3.

We also include a wealth of ground and space-based optical/near-IR data from the COSMOS2020 catalog \citep{Weaver2022}. In brief, this includes \textit{Hubble}/F814W imaging \citep{Koekemoer2007}, the \textit{Spitzer} Cosmic Dawn Survey \citep{EuclidCollaboration2022}, Subaru Telescope Hyper Suprime-Cam (HSC) imaging \citep{Aihara2022}, and UltraVISTA imaging \citep{McCracken2012}. 

The COSMOS field also contains surveys across X-ray, millimeter, and radio wavelengths. Millimeter and submillimeter data includes SCUBA-2 maps at 850\,$\mu$m \citep[S2COSMOS,][]{Simpson2019}, and archival ALMA datasets collated in the A3COSMOS project \citep{Liu2018}. These data are primarily used to refine the dust spectral energy distribution (SED) models and accompanying photo-$z$s. X-ray and radio data are primarily used to discern the presence of powerful active galactic nuclei (AGN), and to quantify whether AGN emission contributes to the 2\,mm flux and/or the galaxy's multiwavelength SED. For these analyses, we use data from the Chandra COSMOS-Legacy program \citep{Civano2016}, the VLA-COSMOS 3\,GHz Large Project \citep{Smolcic2017}, and the MeerKAT MIGHTEE-COSMOS 1.28\,GHz survey \citep{Heywood2022}.

\subsection{NIRCam Photometry}
The NIRCam counterparts to our 2mm-selected sources have resolved morphologies that can be challenging to optimally recover with standard photometric methods using circular or elliptical apertures. We tested measuring multi-wavelength aperture fluxes with \texttt{SE/SE++} \citep{setractor} and found the flux densities to be inconsistent between ground-based (e.g. UltraVISTA) and JWST NIRCam data above 1$\sigma_\mathrm{ground}$. In light of these challenges, we measure fluxes using \texttt{diver} (McKinney et al.~in prep.), an empirically-motivated photometry tool built to preserve the resolution of JWST when measuring PSF-matched multi-wavelength photometry. We now briefly summarize the methodology of \texttt{diver} below.

For each of our sources \texttt{diver} proceeds through the following: First, we make $50^{\prime\prime}\times50^{\prime\prime}$ cutouts around the target in the science image and error/weight image from which a flux will be measured. 
Next, we create a segmentation map using \texttt{photutils} \citep{photutils} from the NIRCam/F444W map, and in some cases include an additional de-blending step to isolate the source from its neighbors. We adopt the arbitrarily shaped segmentation boundary (down to an SNR-per-pixel of 3) of the source as an aperture through which to measure all multi-wavelength photometry. We measure the flux through this aperture in every image and, for lower resolution data (e.g. ground-based and \textit{Spitzer} IRAC data), multiply by a PSF correction factor that accounts for the flux loss due to aperture corrections. We measure flux uncertainties from the local background summed in quadrature with the shot noise, readout noise, and dark current noise all bootstrapped from the error/weight maps using the arbitrarily shaped aperture to faithfully capture the noise scaling per pixel.



The JWST NIRCam (F115W, F150W, F277W, and F444W) fluxes and flux errors measured with \texttt{diver} are generally consistent with the aperture fluxes from \texttt{SE/SE++} within $1\sigma_{\rm diver}$, but not within $1\sigma_{\rm SE/SE++}\ll\sigma_{\rm diver}$. Most notably, the multi-wavelength fluxes from \texttt{diver} are far more consistent between ground-based and JWST data given the expected shape of the rising SEDs into the near-infrared. 

\begin{deluxetable*}{lccccccc}[ht]
\centering
\tablecaption{Ex-MORA $\ge$5$\sigma$ Sources}
 \tablehead{
\colhead{Name} & \colhead{Original Name} & \colhead{R.A.} & \colhead{Dec} & \colhead{S/N$_\mathrm{2\,mm}$} & \colhead{2\,mm Flux Density} & \colhead{z$_\mathrm{spec}^\dagger$} & \colhead{z$_\mathrm{phot}$}\\ 
\colhead{} & \colhead{} & \colhead{(deg)} & \colhead{(deg)} & \colhead{} & \colhead{($\mu$Jy)} & \colhead{} & \colhead{}}
 \startdata
eMORA.0 & MORA-0 & 10:00:15.62 & $+$02:15:49.14 & 19.8 & 1023\,$\pm$\,52 & 4.596 & 4.4\,$_{-0.2}^{+0.2}$ \\
eMORA.1 &  & 09:59:57.28 & $+$02:27:30.54 & 18.8 & 1133\,$\pm$\,60 & 4.618 & 1.0\,$_{-0.1}^{+0.1}$ \\
eMORA.2 & MORA-3 & 10:00:08.05 & $+$02:26:12.24 & 17.2 & 1157\,$\pm$\,67 & 4.630 & 3.4\,$_{-0.1}^{+0.1}$ \\
eMORA.3 &  & 09:59:42.85 & $+$02:29:38.32 & 13.9 & 924\,$\pm$\,66 & 4.340 & 4.4\,$_{-0.1}^{+0.1}$ \\
eMORA.4 & MORA-2 & 10:00:10.15 & $+$02:13:34.74 & 8.8 & 523\,$\pm$\,59 & 3.254$^\mathrm{c}$ & 7.6\,$_{-0.3}^{+0.3}$ \\
eMORA.5$^\mathrm{a}$ & MORA-10 & 10:00:16.58 & $+$02:26:38.04 & 8.2 & 467\,$\pm$\,57 & 2.470 & 7.0\,$_{-0.2}^{+0.5}$ \\
eMORA.6 &  & 09:59:35.34 & $+$02:22:36.51 & 8.2 & 501\,$\pm$\,61 &  & 3.8\,$_{-0.7}^{+0.6}$ \\
eMORA.7 & MORA-1 & 10:00:19.74 & $+$02:32:03.84 & 7.6 & 706\,$\pm$\,93 & 3.340$^\mathrm{c}$ & 4.5\,$_{-0.9}^{+0.7}$ \\
eMORA.8 &  & 09:59:48.89 & $+$02:27:50.93 & 7.3 & 472\,$\pm$\,65 &  & 4.4\,$_{-1.7}^{+1.4}$ \\
eMORA.9 &  & 09:59:33.44 & $+$02:23:47.00 & 7.2 & 444\,$\pm$\,61 &  & 2.5\,$_{-0.2}^{+0.2}$ \\
eMORA.10 &  & 10:00:56.95 & $+$02:20:17.30 & 6.9 & 432\,$\pm$\,63 & 2.494 & 2.6\,$_{-0.2}^{+0.2}$ \\
eMORA.11 & MORA-4 & 10:00:26.37 & $+$02:15:28.14 & 6.6 & 568\,$\pm$\,86 & 5.850 & 5.9\,$_{-2.0}^{+1.2}$ \\
eMORA.12 & MORA-6 & 10:00:28.71 & $+$02:32:03.54 & 6.6 & 621\,$\pm$\,94 & 3.410$^\mathrm{c}$ & 3.4\,$_{-0.1}^{+0.1}$ \\
eMORA.13 & MORA-5 & 10:00:24.15 & $+$02:20:05.64 & 6.4 & 579\,$\pm$\,90 &  & 2.6\,$_{-1.0}^{+1.6}$ \\
eMORA.14 &  & 10:00:33.32 & $+$02:26:02.03 & 6.4 & 505\,$\pm$\,79 & 2.510 & 2.5\,$_{-0.1}^{+0.1}$ \\
eMORA.15$^\mathrm{a}$ &  & 09:59:37.41 & $+$02:23:47.31 & 6.2 & 382\,$\pm$\,61 & 0.741 & 0.7\,$_{-0.1}^{+0.1}$ \\
eMORA.16 &  & 10:01:02.30 & $+$02:22:33.79 & 6.2 & 397\,$\pm$\,64 & 3.717 & 3.4\,$_{-0.1}^{+0.1}$ \\
eMORA.17 &  & 10:01:02.04 & $+$02:28:36.49 & 6.1 & 418\,$\pm$\,68 &  & 3.9\,$_{-1.0}^{+2.5}$ \\
eMORA.18 &  & 09:59:29.22 & $+$02:12:44.30 & 6.1 & 375\,$\pm$\,61 &  & 2.5\,$_{-0.1}^{+0.1}$ \\
eMORA.19 &  & 09:59:38.97 & $+$02:21:26.61 & 6.1 & 370\,$\pm$\,61 &  & 4.2\,$_{-0.3}^{+0.3}$ \\
eMORA.20 &  & 10:00:15.90 & $+$02:24:46.14 & 6.0 & 318\,$\pm$\,53 &  & 5.2\,$_{-0.7}^{+0.7}$ \\
eMORA.21 & MORA-7 & 10:00:11.58 & $+$02:15:05.34 & 6.0 & 370\,$\pm$\,61 &  & 2.9\,$_{-0.3}^{+0.2}$ \\
eMORA.22 &  & 10:01:03.16 & $+$02:21:42.49 & 6.0 & 378\,$\pm$\,64 &  & 4.7\,$_{-0.1}^{+0.1}$ \\
eMORA.23 &  & 09:59:31.16 & $+$02:14:33.50 & 5.9 & 412\,$\pm$\,70 &  & 4.2\,$_{-0.6}^{+0.7}$ \\
eMORA.24 &  & 09:59:57.09 & $+$02:11:36.53 & 5.8 & 354\,$\pm$\,61 & 2.411$^\mathrm{c}$ & 2.3\,$_{-0.2}^{+0.2}$ \\
eMORA.25 &  & 09:59:45.62 & $+$02:11:26.62 & 5.7 & 345\,$\pm$\,61 &  & 4.7\,$_{-0.1}^{+0.1}$ \\
eMORA.26 & MORA-8 & 10:00:25.29 & $+$02:18:46.14 & 5.6 & 491\,$\pm$\,87 & 2.280 & 2.2\,$_{-0.1}^{+0.1}$ \\
eMORA.27 &  & 10:00:47.10 & $+$02:10:16.42 & 5.6 & 379\,$\pm$\,68 & 5.050 & 4.0\,$_{-0.6}^{+0.6}$ \\
eMORA.28 & MORA-9 & 10:00:17.30 & $+$02:27:15.84 & 5.5 & 345\,$\pm$\,63 &  & 4.6\,$_{-1.2}^{+1.3}$ \\
eMORA.29 &  & 10:00:09.83 & $+$02:16:49.74 & 5.4 & 325\,$\pm$\,60 &  & 4.7\,$_{-0.5}^{+1.0}$ \\
eMORA.30 &  & 09:59:29.78 & $+$02:27:39.80 & 5.4 & 348\,$\pm$\,65 &  & 3.1\,$_{-1.3}^{+1.8}$ \\
eMORA.31$^\mathrm{b}$ &  & 09:59:23.28 & $+$02:19:45.18 & 5.3 & 815\,$\pm$\,154 &  & -- \\
eMORA.32 &  & 09:59:43.65 & $+$02:13:40.12 & 5.2 & 345\,$\pm$\,67 &  & 2.6\,$_{-0.2}^{+0.2}$ \\
eMORA.33 &  & 10:00:08.97 & $+$02:20:26.64 & 5.1 & 318\,$\pm$\,62 &  & 2.5\,$_{-0.1}^{+0.2}$ \\
eMORA.34 &  & 10:00:41.34 & $+$02:30:29.93 & 5.1 & 322\,$\pm$\,64 & 0.700 & 0.7\,$_{-0.1}^{+0.1}$ \\
eMORA.35 &  & 09:59:44.03 & $+$02:21:08.92 & 5.0 & 304\,$\pm$\,60 &  & 5.7\,$_{-1.5}^{+1.3}$ \\
eMORA.36$^\mathrm{a}$ &  & 09:59:24.03 & $+$02:27:08.28 & 5.0 & 454\,$\pm$\,91 & 0.710 & 0.7\,$_{-0.1}^{+0.1}$ \\
\enddata  \tablenotetext{}{The $\ge$5$\sigma$ sources identified in the Ex-MORA Survey. Spectroscopic redshifts (if available) are listed in the $z_\mathrm{spec}$ column, while photo-$z$s (estimated via \textsc{bagpipes} or the COSMOS2020 catalog) are listed with their 68\% confidence intervals in the $z_\mathrm{phot}$ column. When available, spectroscopic redshifts are used over photometric redshifts. } \tablenotetext{\dagger}{All spectroscopic redshifts are compiled in Khostovan et al. (in prep), with the original surveys as follows. Rest-frame optical / UV redshifts are from \citet[][]{Lilly2007, Trump2007, Kriek2015, Stott2016, Nanayakkara2016, Marsan2017, Hasinger2018, Wisnioski2019}; Vanderhoof et al. (in prep); Kartaltepe et al. (in prep); and Capak et al. (in prep). CO-derived redshifts are from \citet{Yun2015, Jin2019, Casey2019, Jimenez-Andrade2020, Chen2022}; and Chen et al. (in prep).} \tablenotetext{a}{These sources have significant synchrotron emission without which they would not be detected in this survey.} \tablenotetext{b}{This source is likely spurious, see Section \ref{sec:sample selection}.} \tablenotetext{c}{The spec-$z$s for these sources are based off single CO line detections (with additional supporting information from their photo-$z$ probability distributions, see Appendix Section \ref{sec:degenerate-z}).} 
\end{deluxetable*}  \label{table:sources}

\subsection{Source Selection}

As mentioned in Section \ref{sec:EM survey}, we adopt source positions and flux densities corresponding to the point of their peak signal-to-noise using a ``region grow'' algorithm for peaks at or above 5$\sigma$ significance. Overall, 37 sources are identified in the combined Ex-MORA map at or above 5$\sigma$ significance, with 11 of the original 13 MORA sources recovered \citepalias{Casey2021}. One of the missing original sources (MORA-12) was already believed to be spurious due to lack of detection at any other wavelengths and proximity to the edge of the original maps \citepalias{Casey2021}. In the combined Ex-MORA mosaic, the rms at the edges of the original MORA map -- where MORA-12 was discovered -- are improved. In this new mosaic, MORA-12 disappears entirely, thus confirming its nature as a false object. The other original MORA source (MORA-11) is recovered with an SNR of 4.9$\sigma$ and therefore just misses the 5$\sigma$ threshold for inclusion in this work.\footnote{The extracted flux density for this source from the Ex-MORA map is well within uncertainty of the previous measurement (original $S_\mathrm{2\,mm} = 356\pm69\,\mu$Jy vs. new $S_\mathrm{2\,mm} = 334\pm68\,\mu$Jy), and therefore the discrepancy is not likely an issue with the data reduction and measurement process, but instead likely due to boosting from a coincident noise fluctuation.} We report source coordinates, relevant flux densities, and estimated and/or spectroscopic redshifts for the 37 sources detected at $\ge5\sigma$ in Table \ref{table:sources}. Sources are numbered in inverse order according to their 2\,mm SNRs, with eMORA.0 having the highest SNR across the sample.


Of the 37 sources, 36 ($\sim97\%$) are detected at wavelengths $\lambda_\mathrm{observed}<8\,\mu m$ and 31 ($\sim84\%$) are detected with SCUBA-2 at 850\,$\mu$m. All sources are detected at minimum in the F444W band, with the exception of one source -- eMORA.31. This source is also not detected in any \textit{Spitzer} IRAC bands nor SCUBA-2 imaging, has no ancillary ALMA or radio detections, and is on the edge of the map in a noisy region; thus, we believe this source likely a positive noise fluctuation. We remove eMORA.31 from the analyses in this work, though keep it listed in the Table for posterity. 

In \citetalias{Casey2021}, MORA-10 (herein eMORA.5) had reported synchotron emission significant enough to boost the 2\,mm emission such that without this component, this source would not meet the detection threshold of the survey. Using data from the VLA-COSMOS and MIGHTEE-COSMOS surveys, we find that $23/36$ sources have detections at both the 1.28\,GHz and 3\,GHz frequencies. Using these detections, we calculate synchrotron slopes and extend this to the 2\,mm wavelength. We find that most sources ($n=20$, or $\sim87$\,\%) have estimated sub-percent levels of synchrotron emission in the 2\,mm band. Two out of the three remaining sources are previously identified $z\sim0.7$ AGN \citep[eMORA.15 and eMORA.36,][]{Trump2007}, while the third is the spectroscopically confirmed $z=2.57$ MORA-10/eMORA.5 source previously discussed in \citetalias{Casey2021}, and spectroscopically confirmed in \citet{Kriek2015}. None of these three sources would meet our survey thresholds without significant boosts ($\gtrsim 40\%$ of total 2\,mm flux) from their synchrotron emission. The two $z<1$ radio-boosted sources are also undetected in the SCUBA-2 maps, suggesting that the synchrotron emission is from AGN activity rather than star formation processes. Given that only a small fraction ($n=2/33$, or $<10\%$) of dust-luminous Ex-MORA galaxies lack submillimeter detections, we conclude that 2\,mm follow up on SCUBA-2 sources may prove an efficient filter for high-$z$ dust-obscured systems, as suggested in \citet{Cooper2022} and \citet{Cowie2023}. 


As in \citetalias{Casey2021} and \citetalias{Zavala2021}, we remove the three synchrotron-boosted sources for our statistical analyses of dust continuum emitters, and note that our results presented in Section \ref{sec:results} do not change significantly if these sources were included. 

\subsection{Source Redshifts}

Of the 36 bona fide sources presented herein, 18 have spectroscopic redshifts established in the literature, with 10 at $z_\mathrm{spec}>3$. These spectroscopic redshifts are compiled from several catalogs across multiple wavelengths, with the majority compiled in Khostovan et al. (in prep). This includes the following original surveys: rest-frame optical / UV redshifts are from \citet[][]{Lilly2007, Trump2007, Kriek2015, Stott2016, Nanayakkara2016, Marsan2017, Hasinger2018, Wisnioski2019}; Vanderhoof et al. (in prep); Kartaltepe et al. (in prep); and Capak et al. (in prep); CO-derived redshifts are from \citet{Yun2015, Jin2019, Casey2019, Jimenez-Andrade2020, Chen2022}; and Chen et al. (in prep). A handful of sources have spectroscopic redshifts based on a single CO transition line, meaning their interpretation could be uncertain. We discuss how we address each of these sources individually in Appendix Section \ref{sec:degenerate-z}.

For the majority of the remaining sources, we use photometric redshifts derived via the \textsc{bagpipes} SED-fitting code \citep{Carnall2018} using all of the available UV-to-near-IR data (which now includes JWST data in addition to the data in COSMOS2020). We allow the stellar mass to vary between $10^{6-13}$\,M$_\odot$, and the stellar metallicity from $10^{-3}$ to $10^{0.4}$, both with log-uniform priors. We adopt a non-parametric star-formation history (SFH) model, with five bins in age (fixed bin edges at 10, 30, 100, 300, and 600 Myr) and a constant star-formation history in each bin. We adopt the ``continuity prior,'' to explicitly weigh against significant SFR changes between adjacent bins and avoid overfitting the data \citep[as described in][]{Leja2019}. We include nebular emission, and allow the ionization parameter $\log U$ to vary from --3 to --1. Dust attenuation is included following a \citet{calzetti00} attenuation law with $A_V$ allowed to vary from $0$ to $6$ with a uniform prior. 
For a six sources, \textsc{bagpipes} either failed to produce a fit, or produced an unphysical solution for their photo-$z$. We address these sources individually in Appendix Section \ref{sec:deg-photo-z}. 

We compare the calculated photo-$z$s to spectroscopic redshifts for the 18 available sources in Figure \ref{fig:z_relations} and find good agreement, with a normalized median absolute deviation \citep[NMAD, defined in][]{Brammer2008} of $\sigma_\mathrm{NMAD} = 0.04$. We quantify the catastrophic outlier fraction as the fraction of galaxies with $|z_\mathrm{phot} - z_\mathrm{spec}| \ge (5\times\sigma_\mathrm{NMAD})(1+z_\mathrm{spec})$ \citep[as in e.g., ][]{Weaver2024}, and find that less than a third of the photo-$z$s are 5$\sigma_\mathrm{NMAD}$ away from the one-to-one relation. We also compare newly derived redshifts to those derived in the COSMOS2020 catalog (when available) and find the majority of source redshift estimates are also in similarly good agreement. 

\section{Results \& Discussion} \label{sec:results}
In the following Section, we report statistics on the Ex-MORA sample redshift and 2\,mm flux density distributions. We do not include eMORA.31 (the likely spurious source), nor the three synchrotron emitters from these analyses as the primary goal of this work is to survey thermal dust emission at 2\,mm. Still, we stress that the results agree within 1$\sigma$ uncertainty even if these sources are included.
\begin{figure}
    \centering
    \includegraphics[trim=1cm 0cm 0cm 0cm, width=0.5\textwidth]{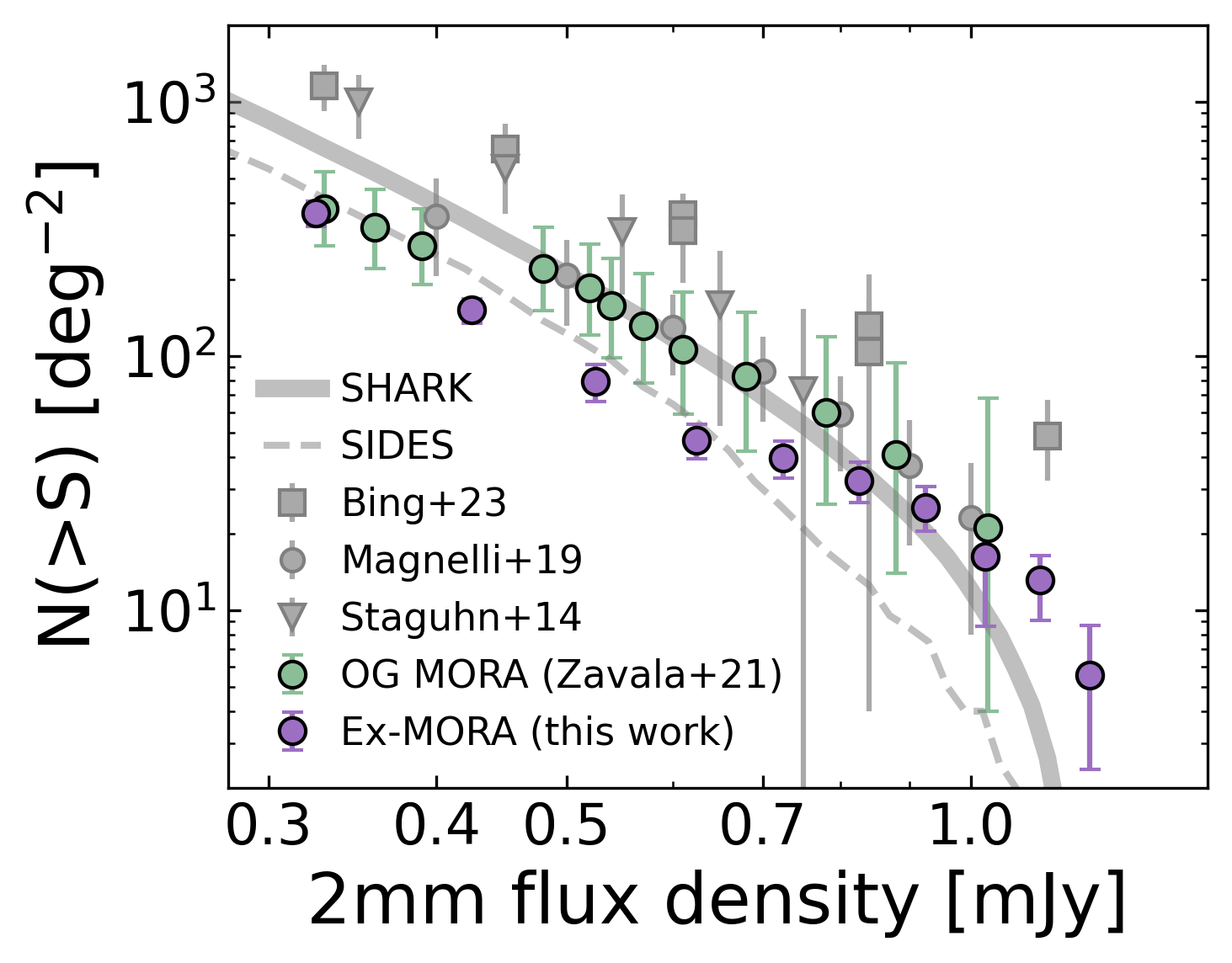}
    \caption{The 2\,mm cumulative number density as a function of solid angle. In grey, we collate data from other 2\,mm blank-field sky surveys  \citep{Staguhn2014,Magnelli2019,Bing2023}, the \textsc{shark} simulation \citep{Lagos2020}, and the \textsc{sides} simulation \citep{Bethermin2017}. Measurements from the original (OG) MORA survey derived in \citetalias{Zavala2021} are presented as light green circles, and measurements from the Ex-MORA Survey (this work) are presented as purple circles. }  
    \label{fig:skydens}
\end{figure}


\subsection{2\,mm Cumulative Number Density} \label{sec:onskydens}
The cumulative number density is defined as the on-sky surface density of sources above a certain flux density threshold, per square degree. We derive the cumulative number density in a similar fashion as presented in \citetalias{Zavala2021} using the 5$\sigma$ source catalog and corresponding completeness and contamination uncertainties, as described at the end of our Section \ref{sec:EM survey}. We describe this process briefly here. Firstly, we estimate the contribution of a source to the cumulative number density through the following equation:
\begin{equation}
    \eta_i(S_i,\sigma_i) = \frac{1-f_\mathrm{cont}(\sigma_i)}{\zeta(S_i)A_\mathrm{eff}}
\end{equation}
for a source with a given deboosted flux density, $S_i$, and SNR, $\sigma_i$; $f_\mathrm{cont}(\sigma_i)$ is the contamination fraction for the measured SNR ($\sigma_i$); $\zeta(S_i)$ is the corresponding completeness for the flux density, and $A_\mathrm{eff}$ is the area of the Ex-MORA map (577\,arcmin$^2$). Cumulative number counts are calculated as $N(>S) = \Sigma_i\eta_i(S_i,\sigma_i)$. 

In order to propagate flux, completeness, and contamination uncertainties into the cumulative number densities, we generate a Monte Carlo simulation. For each iteration, we model individual source flux densities as a Gaussian distribution, setting the dispersion equal to the measured uncertainty. We sample the respective distribution for each individual source and recalculate the SNR given the local noise around that source. We then use this SNR to sample the contamination fraction (which was generated as a function of SNR, as described in Section \ref{sec:EM survey}, and modeled as Gaussians). Each individual source in each iteration also has its own completeness estimates, based on the artificial source injection simulations mentioned in Section \ref{sec:EM survey} and \citetalias{Zavala2021}. We repeat this process $10^3$ times, each time calculating the resulting cumulative number density and corresponding Poisson noise over flux density bins of size 100\,$\mu$Jy. We then calculate the mean, 16th and 84th percentiles, and average Poisson noise in each flux density bin. The resulting cumulative number density is shown in Figure \ref{fig:skydens}, and reported in Table \ref{table:no_dens}. We repeat the above processes while including the four removed sources and find that the differences between cumulative number densities is not significant, and within uncertainty of the `cleaned' cumulative number density. These are also presented in Table \ref{table:no_dens}. 

\begin{figure*}
    \centering
    \includegraphics[width = 0.47\textwidth]{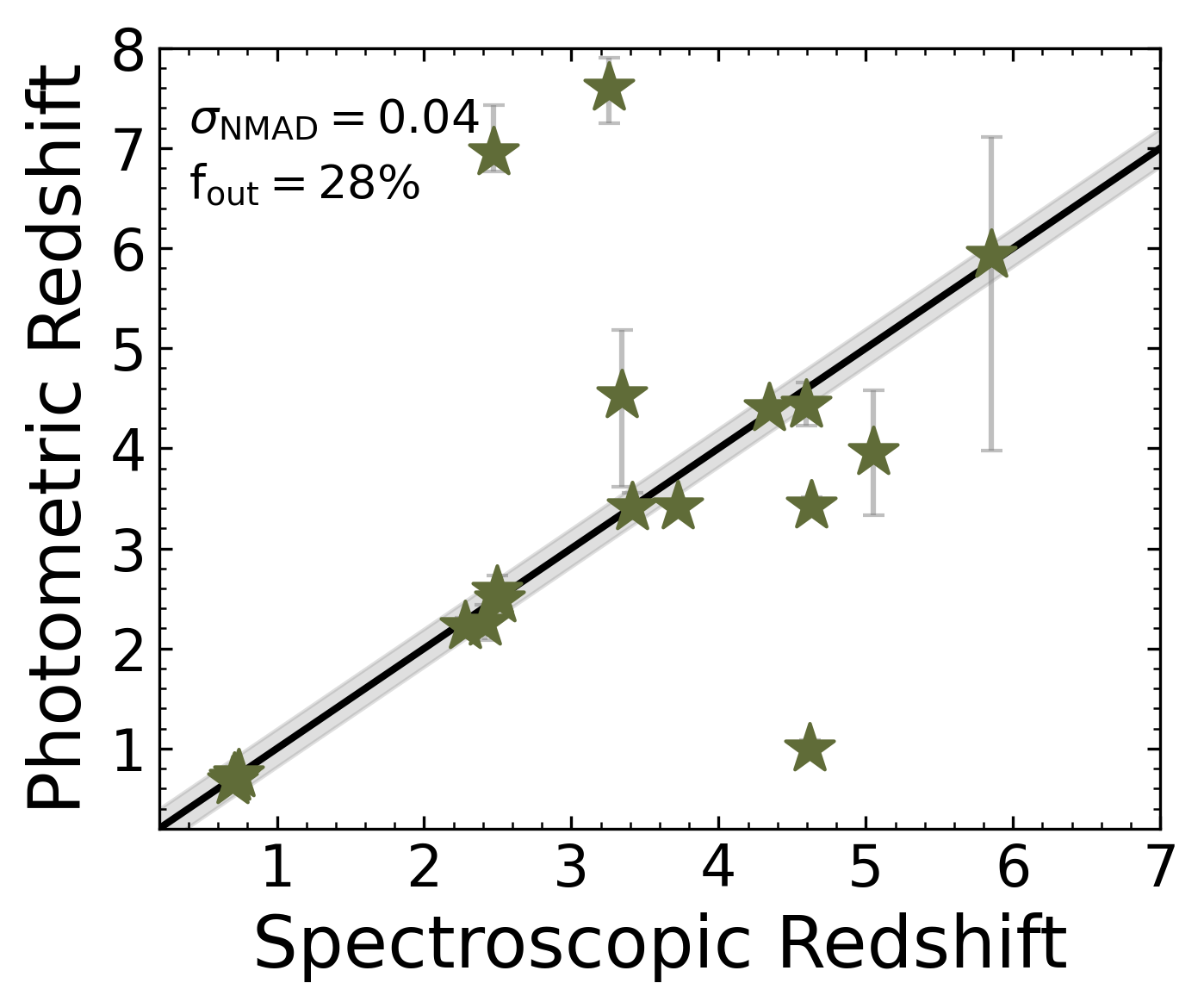}
    \includegraphics[trim = 0cm 0cm 0.5cm 0cm, width = 0.46\textwidth]{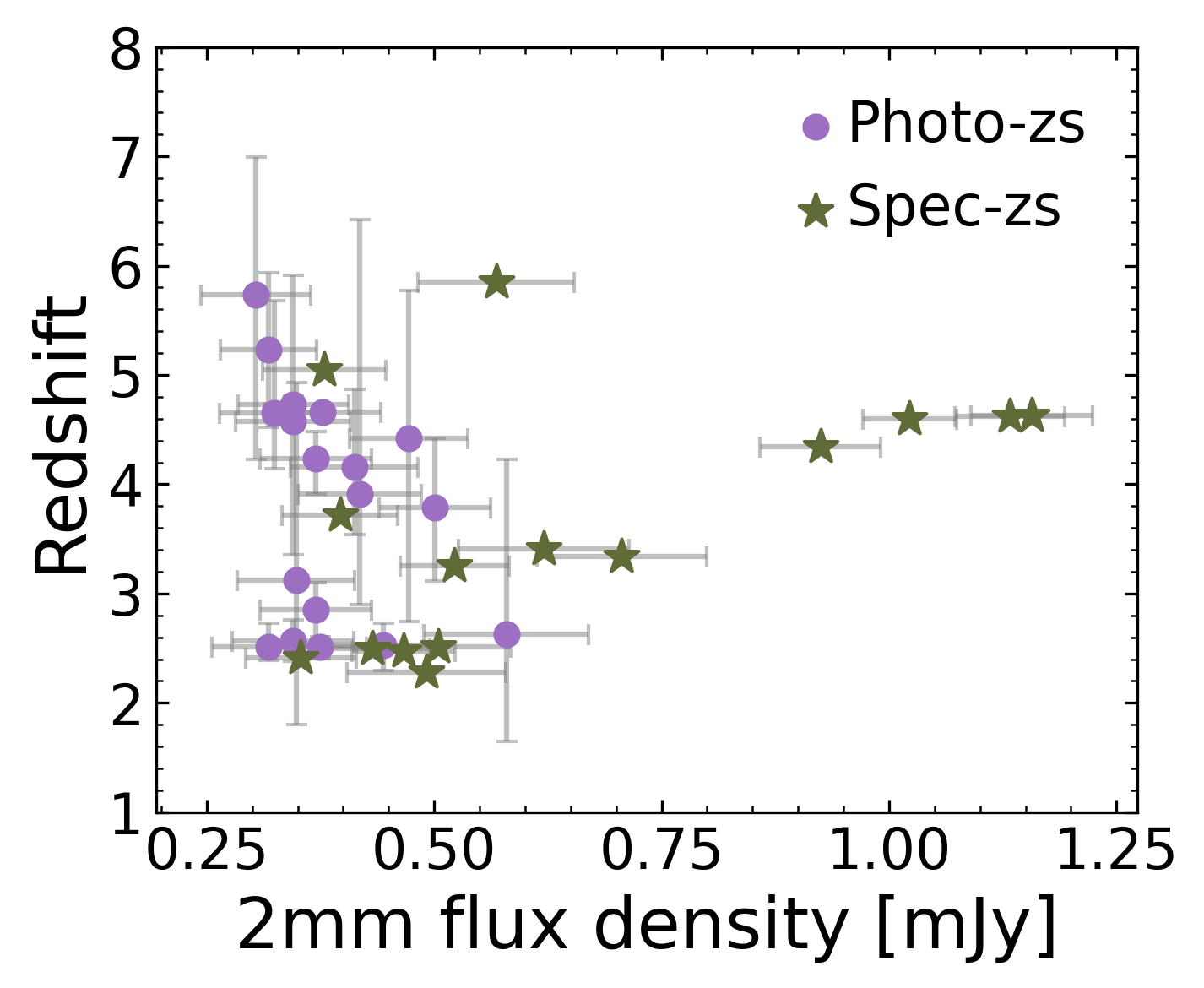}
    \includegraphics[width = 0.48\textwidth]{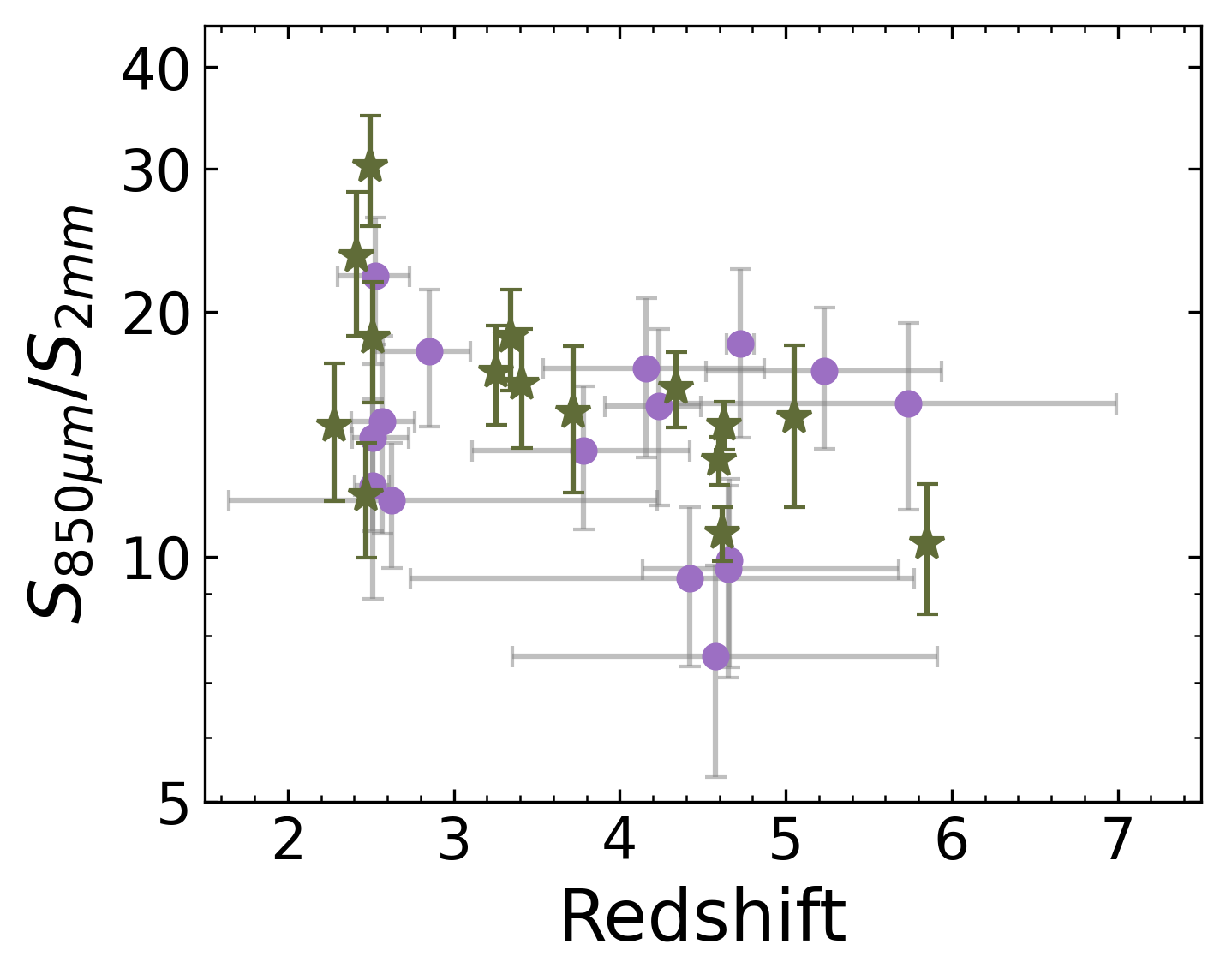}
    \includegraphics[trim = 0cm 0cm 0cm 0cm, width = 0.49\textwidth]{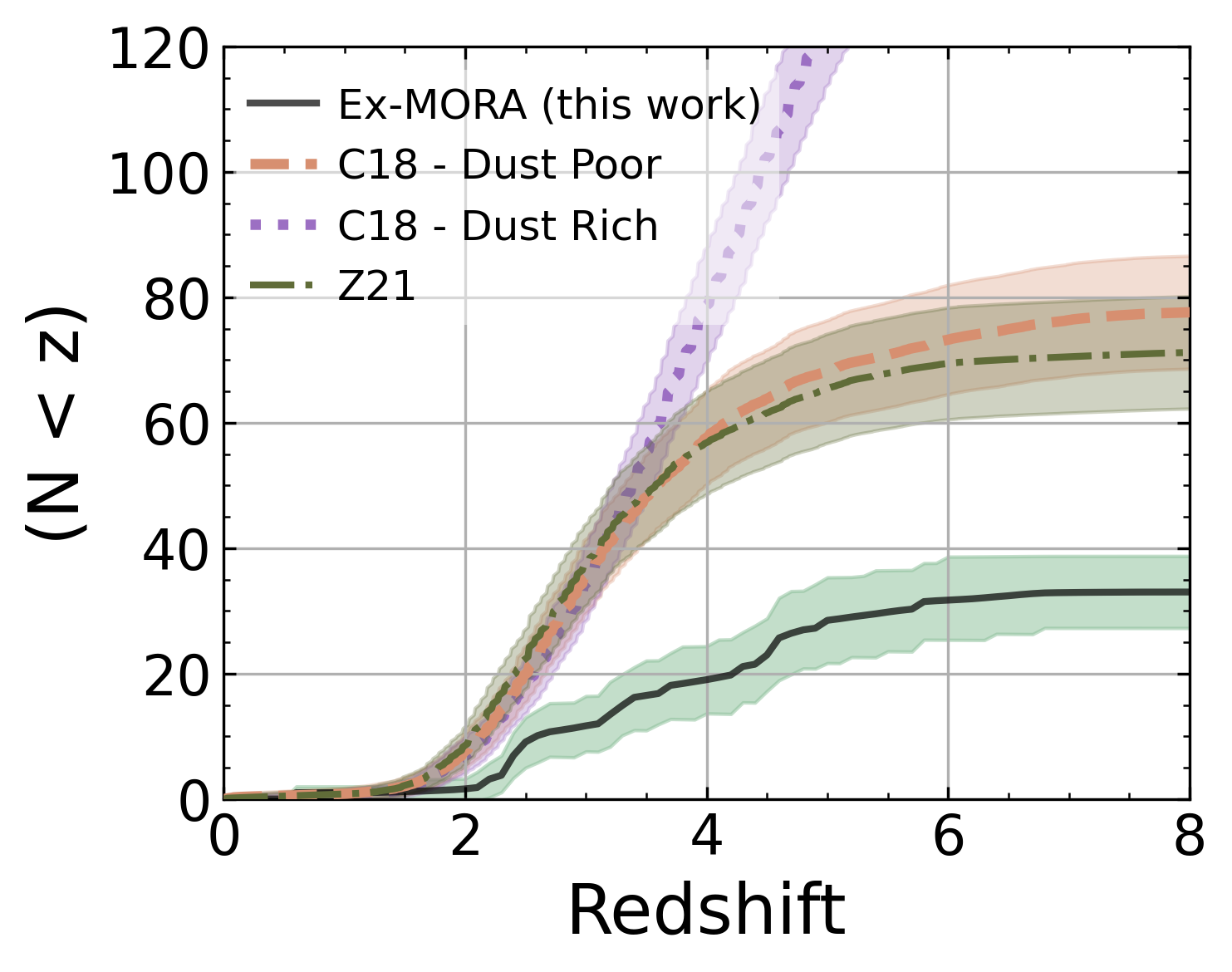}
    \caption{\textit{Top, left:} Comparison between spectroscopic redshifts and photometric redshifts for sources with spec-$z$s from the literature. Solid black line represents the one-to-one relation, the shaded grey region represents the the normalized median absolute deviation (NMAD) of the sample (also reported in the top left), and the outlier fraction (defined in Section \ref{sec:zdist}) is also reported in the top left. \textit{Top, right:} Relationship between 2\,mm flux density and source redshifts. Sources with spectroscopic confirmation are represented as green stars, while sources with only photometric redshifts are shown as purple circles. Some of the spectroscopic sources suggest a positive trend between 2\,mm flux and source redshift, but the trend is washed out by sources with photo-$z$s and a few low-flux $z_\mathrm{spec}\gtrsim4$ sources. \textit{Bottom, left:} Relationship between redshift and $S_\mathrm{850\,\mu m}/S_\mathrm{2\,mm}$ flux ratios, as seen in \citet{Cooper2022} and \citetalias{Casey2021}. \textit{Bottom, right:} The cumulative distribution function of source redshifts is shown in dark green, with light green representing the 1$\sigma$ uncertainty. Results from the original MORA survey \citepalias{Zavala2021} are shown as a green dash-dotted line. CDFs from the \citet{Casey2018a, Casey2018b} theoretical models representing a `dust-poor' versus `dust-rich' early Universe are represented as a pink dashed and purple dotted line, respectively. }
    \label{fig:z_relations}
\end{figure*}

In Figure \ref{fig:skydens}, we compare our results to existing 2\,mm number counts from blank-field surveys across the literature: the 31\,arcmin$^2$ and 250\,arcmin$^2$ GISMO/IRAM surveys reported in \citet{Staguhn2014} and \citet{Magnelli2019}, respectively; the 1010\,arcmin$^2$ COSMOS portion of the IRAM NIKA2 Cosmological Legacy Survey \citep{Bing2023}; the number counts from the original MORA survey in \citetalias{Zavala2021}; and predictions from the \textsc{shark} \citep{Lagos2020} and \textsc{sides} \citep{Bethermin2017} cosmological simulations. On average, all literature estimates are greater than those we calculate by a factor of $2-3$, with the exception of the NIKA2-CLS COSMOS estimates which are $7-8\times$ higher than our estimates. The NIKA2-CLS cumulative number densities shown in Figure \ref{fig:skydens} are the \textit{source} counts, not the galaxy counts. Due to the large beam size of single-dish NIKA2 at 2\,mm ($18$\,''), true galaxy sources are expected to blend together \citep{Bethermin2017}. We apply the ``source-to-galaxy counts correction factor'' derived for this survey in \citet{Bing2023} and find that our estimates are in more agreement, within a factor of $\sim2$, which is similar to the other estimates reported in the literature. Finally, we note that the \textsc{shark} semi-analytical model is roughly successful (within uncertainty) in predicting number densities at $S_\mathrm{2\,mm} \gtrsim 800\,\mu$Jy, but then over-predicts the faint end by $\sim2-3\sigma$; while the \textsc{sides} model performs in an opposite manner, better modeling the observed cumulative number densities at $S_\mathrm{2\,mm} \lesssim 700\,\mu$Jy, but not quite meeting the bright end. We also note that \textsc{shark} produces overall more 2\,mm luminous sources per Mpc$^3$ than \textsc{sides} (and Ex-MORA), which likely produces the systemic offset between the two simulations. As mentioned in \citetalias{Casey2021}, some of the differences in alignment with the observations may driven by intrinsic differences in the redshift distributions of the 2\,mm population. For example, for sources with $S_\mathrm{2\,mm} >= 300$\,$\mu$Jy (like Ex-MORA), both simulations produce average redshifts around $\sim3.2$, which is slightly lower than our survey results ($\sim3.6$). \textsc{sides} predicts that higher redshifts correlate with higher 2\,mm fluxes (which may be supported by the Ex-MORA sample, see Figure \ref{fig:z_relations} and Section \ref{sec:zdist}), and therefore Ex-MORA's slightly higher redshift distribution may drive divergence from the \textsc{sides} model at the bright end. 




We note a distinct difference between the number density shapes / distributions between Ex-MORA and the original MORA survey \citepalias{Zavala2021}. By eye, the MORA cumulative number densities appear to follow a Schechter function, while the Ex-MORA cumulative number densities appear more like a double Schechter function, as Ex-MORA has an apparent comparative deficit in sources with flux densities between $400-800\,\mu$Jy (though the surveys only differ beyond 1$\sigma$ uncertainty at $400-600\,\mu$Jy). We tested whether these differences could be driven by potential variances in data reduction techniques, beam sizes, and/or weather, and find no significant changes in our resulting sample size / cumulative number density. It is possible that completeness corrections for sources at $S_\mathrm{2\,mm} \lesssim 700\,\mu$Jy in the original MORA survey were somewhat overestimated as MORA had overall poorer rms (with only $\sim60\%$ of the map with an rms of $90\,\mu$Jy\,beam$^{-1}$ or less, versus 90\% for the Ex-MORA map; see Fig. \ref{fig:noise}), resulting in low source recovery for fainter sources in the source injection simulations. Ex-MORA, being both deeper and wider, had only significant completeness corrections for sources with flux densities $\lesssim 450\,\mu$Jy, which corresponds to the $5\sigma$ threshold for a map with rms ($\sigma$) $=90\,\mu$Jy\,beam$^{-1}$. For MORA, the corresponding 90$\%$ rms is $\sim156\,\mu$Jy\,beam$^{-1}$, setting the 5$\sigma$ threshold at 780\,$\mu$Jy, which is near where we begin to see the departure between survey results as the completeness corrections in MORA become significant ($>10\%$). Both surveys estimate the same number of sources at the faintest end (within uncertainty), where $S_\mathrm{2\,mm} \sim 300\,\mu$Jy, which at $5\sigma$ corresponds to an rms of 60\,$\mu$Jy\,beam$^{-1}$. While the rms varies over both maps and Ex-MORA is generally deeper, both Ex-MORA and MORA have a small fraction of the map at or below an rms of 60\,$\mu$Jy\,beam$^{-1}$ ($\sim2-6\%$), which means that the completeness corrections for sources near this flux density are likely similar across both surveys. 


\begin{deluxetable}{c | ccc | ccc}[t!]
\centering
\tablecaption{Ex-MORA 2\,mm Cumulative Number Densities}
 \tablehead{
 \colhead{} & \multicolumn3c{All Sources} & \multicolumn3c{Cleaned Sample} \\
\colhead{$S_\mathrm{2\,mm}$} & \colhead{$N(>S)$} & \colhead{$-\delta\,N$} &  \colhead{$+\delta\,N$} & \colhead{$N(>S)$} & \colhead{$-\delta\,N$} &  \colhead{$+\delta\,N$} \\ \colhead{(mJy)} & \colhead{(deg$^{-2}$)} & \colhead{(deg$^{-2}$)} & \colhead{(deg$^{-2}$)} & \colhead{(deg$^{-2}$)} & \colhead{(deg$^{-2}$)} & \colhead{(deg$^{-2}$)}
}
 \startdata
0.30 & 365 & 42 & 41 & 325 & 36 & 38 \\
0.40 & 152 & 17 & 16 & 134 & 16 & 15 \\
0.50 & 79 & 13 & 13 & 72 & 13 & 12 \\
0.60 & 47 & 7 & 7 & 39 & 7 & 6 \\
0.70 & 40 & 7 & 7 & 32 & 6 & 6 \\
0.80 & 32 & 6 & 6 & 25 & 5 & 5 \\
0.90 & 25 & 5 & 5 & 25 & 5 & 5 \\
1.00 & 16 & 8 & 7 & 16 & 8 & 7 \\
1.10 & 13 & 4 & 3 & 13 & 4 & 3 \\
1.20 & 6 & 3 & 3 & 6 & 3 & 3 \\
 \enddata \tablenotetext{}{The `Cleaned Sample' corresponds to Figure \ref{fig:skydens} and represents estimates with the spurious source (eMORA.31) and significant synchrotron emitters (eMORA.5, eMORA.15, and eMORA.36) removed. }
\end{deluxetable} \label{table:no_dens}

\subsection{Redshift Distribution of 2\,mm Sources} \label{sec:zdist}
In Figure \ref{fig:z_relations}, we present the cumulative distribution function (CDF) of Ex-MORA sources. We calculate the CDF over 10$^3$ iterations. For each instance, we sample from each source's redshift probability distribution function, \textit{p(z)}, for galaxies with photo-$z$s, while galaxies with spec-$z$s are treated as delta functions. We then calculate the mean CDF and 68\% confidence interval from the 10$^3$ CDFs generated in the 10$^3$ iterations. 

Based on the CDF, we find that 95$^{+2}_{-1}$\% of 2\,mm selected sources are at $z>2$, 66$^{+3}_{-3}$\% at $z>3$, and 43$^{+4}_{-4}$\% at $z>4$. This means that most 2\,mm selected sources are likely at $z>3$, and nearly \textit{half} are likely at $z>4$. We sample from the CDF and corresponding uncertainties to measure a median redshift of $\langle z \rangle = 3.6^{+0.1}_{-0.2}$ which is remarkably equivalent to the estimate derived in \citetalias{Casey2021} over a smaller area using less sensitive OIR data. This average is higher than the smaller and shallower GISMO/IRAM 2\,mm Survey in \citet[][$z\sim3$]{Staguhn2014}, but is in more agreement with the average found in the corresponding follow up survey that is $\sim8\times$ larger in area than the original \citep{Magnelli2019}, with $\langle z \rangle  = 3.9$. 

In Figure \ref{fig:z_relations}, we also show redshift CDFs from the \citet{Casey2018a, Casey2018b} theoretical models representing a `dust-poor' versus `dust-rich' early Universe. In this scenario, a `dust-poor' early Universe is one without intense, dusty starbursts (with obscured SFRs $> 100$\,M$_\odot$\,yr$^{-1}$), \textit{not} an early Universe devoid of galaxies with significant fractions of obscured star-formation \citep[see e.g.][]{fudamoto20, Algera2023}. The models are generated by sampling over a simulated 2\,mm survey of the same on-sky area as Ex-MORA. While the Ex-MORA survey is successful at identifying primarily $z>3$ IR luminous galaxies, the results are still consistent with -- and even markedly below -- the `dust-poor' model. This means that the number density of heavily obscured, starbursting galaxies at high-$z$ should decrease with increasing redshift at $z>3$, and that the cosmic star formation rate density may be less dominated by extreme systems of obscured star formation during these epochs. The former implication is demonstrated in Section \ref{sec:nodens}, while the latter will be presented in a forthcoming paper. Combining cumulative number densities from the original MORA survey with 1.2\,mm and 3\,mm number counts from the literature, \citetalias{Zavala2021} generated an independent prediction for the cumulative redshift distribution of 2\,mm selected sources\footnote{We note that, while \citetalias{Zavala2021} uses MORA number counts to generate their predictions, the combination of other wavelengths and the methods employed therein make their predictions independent from those derived from direct source characterization, as done in this work. See \citetalias{Casey2021} for more details.}, predicting roughly twice the number of $z>3$ sources than what we measure, even though their predictions are also consistent with a `dust-poor' early Universe. They estimated that the evolution of the characteristic number density of IR luminous galaxies $\Phi_\star \propto (1+z)^{\psi}$ is steeply declining at $z\gtrsim2$, with $\psi = -6.5^{+0.8}_{-1.8}$. Given that Ex-MORA produces even fewer 2\,mm luminous galaxies at $z>3$ than expected based on the model in \citetalias{Zavala2021}, it is likely that $\Phi_\star$ is declining at an even more rapid rate, with $\psi < -6.5$. Future work using Ex-MORA to constrain the evolution of the IR luminosity function parameters at higher detail will be forthcoming. 

We also explore the relationship between source redshifts and their 2\,mm flux densities, as well as their ratios between 850\,$\mu$m and 2\,mm flux densities, in Figure \ref{fig:z_relations}. This latter ratio was shown to potentially correlate with source redshifts such that lower ratios correspond to higher redshifts as the 850\,$\mu$m and 2\,mm bands begin mutually probing the peak of dust emission \citep{Casey2021, Cooper2022}. As seen in Casey et al. and Cooper et al., both relationships show potential for correlation such that higher redshift sources are on average brighter at 2\,mm, and have lower $S_\mathrm{850\,\mu m}/S_\mathrm{2\,mm}$ ratios. This appears particularly true for most of the spectroscopically confirmed sources in the Ex-MORA sample. However, when considering sources with photometrically derived redshifts, both of these trends appear to wash out. Future followup spectroscopic surveys over larger samples will prove fruitful in ascertaining whether these values are truly correlated with galaxy redshifts at $z\gtrsim3$. 

\subsection{Volume Density of 2\,mm Sources at $z\gtrsim3$}  \label{sec:nodens}
\begin{figure}
    \centering
    \includegraphics[trim = 1cm 0cm 0cm 0cm, width=0.49\textwidth]{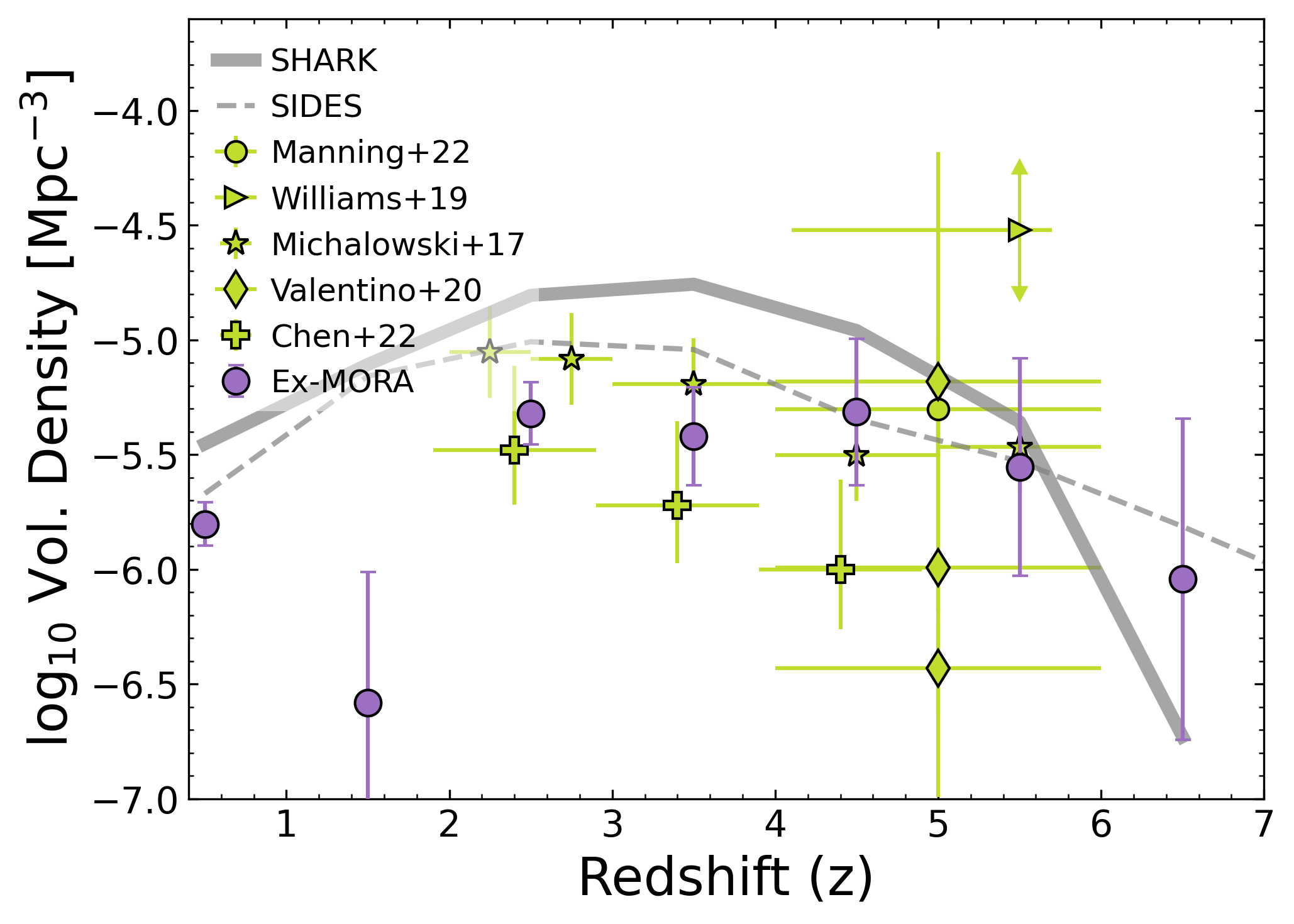}
    \caption{The estimated volume density evolution of 2\,mm sources in the Ex-MORA Survey is shown as purple dots, with corresponding values listed in Table \ref{table:voldens}. Comparative studies on DSFG number densities using a variety of other selection techniques / wavelengths are delineated in green \citep{michalowski17,Williams2019, Valentino2020, Chen2022, Manning2022}. We also show estimates of 2\,mm-bright sources ($S_\mathrm{2\,mm} >= 0.3$\,mJy) from the \textsc{shark} \citep{Lagos2020} and \textsc{sides} \citep{Bethermin2017} in grey. }
    \label{fig:vol_dens}
\end{figure}

The large contiguous area of the Ex-MORA survey, combined with rich ancillary data (including JWST imaging) and a well defined flux threshold, enables strong constraints on high-$z$ DSFG volume densities. In Figure \ref{fig:vol_dens}, we present the volume density of Ex-MORA sources over $z=0-8$ alongside estimates collated from the literature using a variety of DSFG samples \citep{michalowski17, Williams2019, Lagos2020, Valentino2020, Manning2022, Chen2022}. We calculate our estimates using the same method presented in Section \ref{sec:zdist} in volumes encompassed by redshift bins of size $\delta z = 1$ over the Ex-MORA survey area. We add in quadrature cosmic variance, $\sigma_\mathrm{CV}$, over the various redshift bins following the prescription in McKinney et al. (in prep) using the standard assumptions and cookbook presented in \citet{Moster2011}. We note that $\sigma_\mathrm{CV}$ dominates the volume density uncertainties for 2\,mm-luminous sources at $z>2$. The resulting calculations and corresponding uncertainties are presented in Table \ref{table:voldens}. 

Current volume density estimates of submillimeter-luminous DSFGs at $z\gtrsim3$ are in tension by a factor of $\sim100$, ranging from $\sim 3 \times 10^{-5}$\,Mpc$^{-3}$ \citep{Williams2019} to $\sim 4 \times 10^{-7}$\,Mpc$^{-3}$ \citep{Valentino2020, Valentino2023}. In this sample, we derive a roughly steady volume density of 2\,mm -bright galaxies from $z=2$ to 5 of $\sim4-5 \times 10^{-6}$\,Mpc$^{-3}$. This estimate agrees well with estimates derived from hundreds of SCUBA-2 selected DSFGs in \citet{michalowski17}, but is slightly higher (specifically at $z>3$) than the (duty cycle corrected) volume densities reported in \citet{Chen2022}, though the latter study at present has a much smaller sample size. We also compare to the \textsc{shark} \citep{Lagos2020} and \textsc{sides} \citep{Bethermin2017} cosmological models by only including sources with $S_\mathrm{2\,mm} > 300$\,$\mu$Jy (the lower flux density threshold of our sample). Despite general success in modeling the bright end of the 2\,mm cumulative number density, both models over predict the abundance of 2\,mm sources across $z\sim2-4$ by roughly a factor of 3; however, the models do exhibit a similar rate of population decrease from $z=4$ to $z=7$, and are in more agreement with the observed number densities at $z>4$. 

\begin{deluxetable}{ccccc}[ht]
\centering
\tablecaption{Ex-MORA 2\,mm Volume Densities}
 \tablehead{
\colhead{Redshift} & \colhead{log$_\mathrm{10}$($n$/Mpc$^{-3}$)} & \colhead{$\sigma_\mathrm{CV}$} 
}
 \startdata
$0< z <1$ & $-5.80 \pm 0.03$ & 0.09\\
$1< z <2$ & $-6.57 \pm 0.56$ & 0.09\\
$2< z <3$ & $-5.32 \pm 0.04$ & 0.13\\
$3< z <4$ & $-5.42 \pm 0.07$ & 0.20\\
$4< z <5$ & $-5.32 \pm 0.07$ & 0.31\\
$5< z <6$ & $-5.55 \pm 0.12$ & 0.46\\
$6< z <7$ & $-6.04 \pm 0.28$ & 0.64\\
\enddata 
\end{deluxetable} \label{table:voldens}

\subsection{Other High-$z$ Red Selection Techniques}
In recent years, several new techniques were developed to efficiently select $z>3$ dusty galaxies, with a special focus on sources with faint or no emission at optical/near-IR (OIR) wavelengths. These ``OIR-dark'' galaxies are believed to live predominantly at high redshifts ($z>3$), contribute significantly to the cosmic star formation rate density ($\sim 10-50\%$), and serve as a primary parent population for the first massive quiescent galaxies at $z\sim3-5$ \citep{Wang2019, Talia2021, Manning2022, Enia2022, Barrufet2023, Xiao2023b, Xiao2023, vanderVlugt2023}. It is important to note, however, that each of these methods selects different populations of dusty galaxies \citep[e.g. Section 5.2 in][]{Talia2021}, with variations over average star formation rates, stellar masses, and redshifts. In the following, we apply some of these techniques to the Ex-MORA sample to understand how this population may or may not be included in such studies. 

We first apply the radio\,$+$\,OIR-dark selection technique defined in \citet{Gentile2024}. This selection technique requires that galaxies have 3\,GHz flux densities $>12.65\,\mu$Jy and do not have a detection in COSMOS2020 \citep{Weaver2022}; the reddest band used for source detection in COSMOS2020 is the UltraVISTA $K_s$ filter ($\lambda \sim 2.2$\,$\mu$m) with a 3$\sigma$ limiting depth at m$_\mathrm{AB} \sim 25.5$. We find that $22/36$ ($\sim60\%$) sources have sufficiently bright 3\,GHz emission ($>12.65$\,$\mu$Jy) to pass the first selection criterion, though three sources are known AGN and thus should be removed from this sample. Out of the remaining 19 Ex-MORA galaxies, only three sources are sufficiently faint to have no counterpart in the COSMOS2020 catalog. If we instead use the $F277W$ band as a proxy for the $K_s$-band (with thee 3$\sigma$ magnitude limit mentioned above), we only recover one source with the other two becoming slightly too bright to be captured. 

We next apply the OIR-dark selection technique defined in \citet{Wang2019} but updated using JWST NIRCam filters \citep{Barrufet2023, Gottumukkala2024}. The OIR-dark selection technique, when used with JWST bands, requires F150W$ - $F444W\, $> 2.1$ and a faint or non-detection in the F150W band ($m_{F150W} > 25$). We find that a little under half ($n=14/33$) of Ex-MORA sources would be captured with this technique. Of the sources not captured: most have sufficiently red F150W$ - $F444W colors to meet the first criterion (F150W$ - $F444W $> 2.1$), but their F150W magnitudes are too bright ($< 25$\,mag). We note that \textit{all} of the sources captured by this technique are undetected in the F150W band; given the observational depths of the F150W band in the COSMOS-Web survey \citep[m$_\mathrm{AB} \lesssim 27$, ][]{cosmos-web}, it is likely that that these sources are truly ``dark''. Sources without $F150W$ detections have an average redshift of $z\sim 4.4$, while those with $F150W$ detections sit near $z\sim 3.3$ (though these averages are at the edge of 1$\sigma$ uncertainty of one another). 

Combined, it seems that these opt/NIR and radio selection techniques capture roughly $10-40$\% of all 2\,mm-bright sources, supporting the idea that the heavily dust obscured population is quite heterogeneous. The population densities derived using these methods are similar to our estimates when considering all Ex-MORA sources at $z>3$ \citep[$\sim1-3 \times 10^{-5}$\,Mpc$^{-3}$, ][]{Talia2021, Xiao2023}, yet there is little to no overlap between sources identified via these three methods. This is particularly troubling as it implies that up to two-thirds of the $z>3$ dusty galaxy population is missed when using only one of these selection techniques \citep[and there may be more dust covered SFR in galaxies at these epochs than previously thought, e.g. ][]{Algera2023, Zimmerman2024}. If confirmed, the consequences of such biases are wide-sweeping, affecting theories and estimates on early and rapid metal enrichment of the ISM, the dust-obscured component of the cosmic star formation rate density, and early massive quiescent galaxy progenitor populations. This also implies that compiling the most complete samples of dusty, star forming galaxies at high-$z$ requires a combination of observations across several energy regimes, further solidifying how synergies between major observing facilities such as JWST and ALMA are critical to this aim. 




\section{Summary \& Conclusions} \label{sec:conclusions}
We have presented the Ex-MORA Survey, the largest ALMA blank-field map covering 577\,arcmin$^2$ at 2\,mm. This work is an expansion on the original MORA survey presented in \citetalias{Zavala2021} and \citetalias{Casey2021}. We detect 37 sources above $5\sigma$ significance, one of which is likely a false-positive, and three of which likely have 2\,mm fluxes boosted significantly by synchrotron emission, leaving a total of 33 robust detections from predominantly thermal dust emission. All bona fide sources are detected at minimum in the NIRCam $F444W$ band. 

Through a combination of spectroscopic and photometric redshifts spanning $z\sim2-6$, we estimate a median redshift of $\langle z \rangle  = 3.6^{+0.1}_{-0.2}$. Two-thirds of the sample are estimated at $z>3$, and just under half of the sample is estimated at $z>4$. We estimate the volume density of Ex-MORA sources and measure a decrease in the population from $z\sim4$ to $z\sim7$, with an integrated volume density of $\sim1-3\times10^{-5}$\,Mpc$^{-3}$ at $z>3$. In upcoming work, we will present the physical properties of individual Ex-MORA sources and explore them in the context of galaxy evolution paradigms. 

While the Ex-MORA survey is successful at identifying a strong sample of primarily $z>3$ IR luminous galaxies, the results are still consistent with the `dust poor' early Universe model presented in \citet{Casey2018a}. At face value, this implies that searches for high-$z$ heavily dust-obscured galaxies will become more difficult with increasing redshift as these sources are incredibly rare \citep[though known to exist e.g., ][]{Zavala2018Nature, Marrone2018, Akins2023, Hygate2023}. In this work, we find that other techniques developed to discover $z>3$ dusty galaxies \citep[e.g.,][]{Wang2019, Talia2021, Barrufet2023, Gottumukkala2024} capture as much as 40\% and as little as 10\% of the Ex-MORA galaxy sample. These findings support the idea that multiple modes of selection across multiple wavelengths and energy regimes are necessary for a complete census of the high-$z$ dusty galaxy population, and that synergistic efforts with observatories such as JWST, ALMA, and the VLA will continue to be critical to these efforts. 


\section{Acknowledgements}

We honor the invaluable labor of the maintenance and clerical staff at our institutions, whose contributions make our scientific discoveries a reality. This work was developed and written in central Texas on the lands of the Tonkawa, Comanche, and Apache people. Learn more about the history and present life of Austin's Indigeous peoples at \href{https://www.kut.org/texas/2022-03-31/where-have-austins-indigenous-people-gone}{https://www.kut.org/texas/2022-03-31/where-have-austins-indigenous-people-gone}.

ASL thanks Charlie and Patrick Long for the love, support, and precious moments baking in the sun together. ASL would also like to thank the good people at NRAO Charlottesville for their help in staging and reducing this data, for sharing remote computing resources, for the ALMA Ambassador Program, and for the Visitor Support Program. Specifically, thank you to G. Privon, S. Wood, and L. Barcos-Munoz. ASL would also like to thank the LUMA, VanguardSTEM, Black in Astro, and AAS CSMA organizations for providing community and support, with special thanks to A. Harriot, G. Ezeka, M. Sgouros, K. Klass, R. McNair, L. Valencia, A. Walker, N. Cabrera Salazar, and J. Gonzalez Quiles. 

ASL and SMM acknowledge support for this work provided by NASA through the NASA Hubble Fellowship Program grant \#HST-HF2-51511 and \#HST-HF2-51484, awarded by the Space Telescope Science Institute, which is operated by the Association of Universities for Research in Astronomy, Inc., for NASA, under contract NAS5-26555. JH acknowledges support from the ERC Consolidator Grant 101088676 (``VOYAJ''). JAZ acknowledges funding from JSPS KAKENHI grant number KG23K13150. FG acknowledges the support from grant PRIN MIUR 2017-20173ML3WW\_001 `Opening the ALMA window on the cosmic evolution of gas, stars, and supermassive black holes.' ET acknowledges support from ANID Basal program FB210003 (CATA) and FONDECYT Regular 1190818 and 1200495. JBC acknowledges funding from the JWST Arizona/Steward Postdoc in Early galaxies and Reionization (JASPER) Scholar contract at the University of Arizona. AWSM acknowledges the support of the Natural Sciences and Engineering Research Council of Canada (NSERC) through grant reference number RGPIN-2021-03046. MA acknowledges support from the ANID BASAL project FB210003

The National Radio Astronomy Observatory is a facility of the National Science Foundation operated under cooperative agreement by Associated Universities, Inc. ALMA is a partnership of ESO (representing its member states), NSF (USA) and NINS (Japan), together with NRC (Canada), MOST and ASIAA (Taiwan), and KASI (Republic of Korea), in cooperation with the Republic of Chile. The Joint ALMA Observatory is operated by ESO, AUI/NRAO and NAOJ. This work is based [in part] on observations made with the NASA/ESA/CSA JWST. The data were obtained from the Mikulski Archive for Space Telescopes at the Space Telescope Science Institute, which is operated by the Association of Universities for Research in Astronomy, Inc., under NASA contract NAS 5-03127 for JWST. These observations are associated with program JWST \#1727 and \#1837. Some of the data presented in this paper are available on the Mikulski Archive for Space Telescopes (MAST) at the Space Telescope Science Institute. The specific observations can be accessed via doi:10.17909/qhb4-fy92 and doi:10.17909/T94S3X. This work made use of Astropy:\footnote{http://www.astropy.org} a community-developed core Python package and an ecosystem of tools and resources for astronomy \citep{astropy:2013, astropy:2018, astropy:2022}. 

\bibliography{references}{}
\bibliographystyle{aasjournal}

\appendix

\setcounter{table}{0}
\renewcommand{\thetable}{A\arabic{table}}

\section{Individual Source Redshift Solutions}
\subsection{Approaching Degenerate Source Spectroscopic Redshifts} \label{sec:degenerate-z}

A handful of Ex-MORA sources have spectroscopic redshifts generated from a single CO emission line. Without a second line detection (from e.g. a higher or lower $J-$transition, [C I] or [C II]), interpretation of the corresponding spectroscopic redshift is degenerate across discrete redshift values \citep[see e.g.,][]{Jin2024}. We discuss the handful of sources for which this degeneracy is possible, and how we decided which redshift to report.

For two objects (eMORA.7 and eMORA.12), the detected line is attributed to a CO (3-2) transition line, placing these sources at $z \sim 2.31$ and 2.26, respectively. However, given the spectral coverage of the observations (ALMA band 3 only, PID:2019.1.01600.S and 2021.1.00246.S), it is possible to have a single line detection also at CO (4-3), which would place these objects instead at $z \sim 3.34$ and 3.41, respectively. Indeed, the photometric redshifts developed in this work support a $z>3$ solution for both of these sources, and agrees with the CO(4-3) redshift within 1$\sigma$ uncertainty and $ \int_{3}^{\infty}$\,p(z)\,$>85\%$. Thus, for these two sources, we adopt the $z\sim3.4$ solutions in this work. 

Another source (eMORA.24) is similar to the above two sources (a single detection reported as CO(3-2) and a corresponding $z<3$ redshift estimate in Chen et al. (in prep)), however the photo-$z$ estimates strongly agree with the predicted spectroscopic redshift at $z=2.411$ within 1$\sigma$ (and $\int_{3}^{\infty}$\,p(z)\,$<5\%$). For this source, we adopt the reported spectroscopic redshift of 2.411. 

Finally, one source (eMORA.4) has a single line detection attributed to CO(4-3) with a corresponding redshift of $z=3.254$. The estimated photo-$z$ from this work is high, at $z\sim7.6$. If true, this would require a CO(8-7) emission line \textit{and} have a much higher likelihood of detecting multiple $J$-transition lines because (a) the distance between rest-frame wavelengths of CO emission lines becomes smaller with increasing $J$-transitions; and (b) emission line brightness increases with increasing $J$-transitions. Thus, we reject the high-$z$ photometrically-derived solution and adopt its redshift as $z=3.254$. 

\subsection{Photometric Redshifts Solutions for Unique Sources} \label{sec:deg-photo-z}

Three sources (eMORA.6, eMORA.20, and eMORA.28) are estimated to live at $z\sim1.5$, according to the \textsc{bagpipes} fit. If true, this would make these sources near HyLIRGS (L$_\mathrm{IR} \sim 10^{13}$\,M$_\odot$) and therefore significantly detected in \textit{Herschel} SPIRE bands, with $S_\mathrm{850\,\mu m} \gtrsim 50$\,mJy \citep{casey2014dusty}. All three sources are not detected above 1$\sigma$ significance in the existing \textit{Herschel} maps \citep{oliver2012herschel}, and therefore we reject these solutions. For eMORA.6 and eMORA.28, we adopt the estimate reported in the COSMOS2020 catalog, which sits at $z>3$ for both sources. eMORA.20 was not captured in the original COSMOS2020 catalog as it is faint / invisible at $\lambda_\mathrm{obs} < 2$\,$\mu$m (though still visible in the K$_s$-band). Therefore it does not have a redshift estimate in COSMOS2020. For this source, we instead run the \textsc{cigale} energy-balance SED and photo-$z$ fitting software with a parameterization following \citet{Long2024} and find a solution of $z_\mathrm{phot}\sim5.5$. 

Finally, three sources are estimated at $z\sim7$ according to their \textsc{bagpipes} fits (eMORA.21, eMORA.23, and eMORA.32). However, upon inspection, two of these sources (eMORA.21 and eMORA.32) have large effective radii in their $F444W$ images ($r_e \sim 1-2''$). This would correspond to physical sizes on the order of $\sim5-11$\,kpc at $z\sim7$, which is significantly larger ($2-10\times$) than the measured sizes of star forming galaxies at such early epochs \citep{Bowler2017}, and more in agreement with the sizes of dusty galaxies around Cosmic Noon \citep[though still slightly larger, e.g. ][]{Burnham2021}. \textsc{bagpipes} completely failed to fit eMORA.23, despite outputting a redshift estimate and corresponding galaxy properties. For all three sources, we adopt the estimates reported in the COSMOS2020 catalog, which sit at $z>2$. 

\end{document}